\documentclass[pre,twocolumn,showpacs,amsmath,amssymb]{revtex4} 
 
\usepackage[english]{babel}
\usepackage{epsfig}

\newcommand{\me}{e}
\newcommand{\dd}{d}
\newcommand{\eps}{\varepsilon}
\newcommand{\sgn}{\mathrm{sgn}}

\newcommand{\In}{I}
\newcommand{\Hn}{H}
\newcommand{\ivec}{\ensuremath{\boldsymbol{i}}}
\newcommand{\jvec}{\ensuremath{\boldsymbol{j}}}
\newcommand{\bsig}{\boldsymbol{\sigma}}

\newcommand{\dt}{\ensuremath{\Delta t}} 
\newcommand{\tw}{\ensuremath{t_\mathrm{w}}} 
\newcommand{\Cst}{C_\mathrm{st}}
\newcommand{\cst}{c_\mathrm{st}}
\newcommand{\Cag}{C_\mathrm{ag}}

\newcommand{\Rst}{R_\mathrm{st}}
\newcommand{\rst}{r_\mathrm{st}}
\newcommand{\Rag}{R_\mathrm{ag}}

\newcommand{\Sst}{\chi_\mathrm{st}}
\newcommand{\Sag}{\chi_\mathrm{ag}}

\newcommand{\eq}{\mathrm{eq}}
\newcommand{\taueq}{\tau_{\mathrm{eq}}}

\begin{document} 

\title{Observable Dependent Quasi-Equilibrium in Slow Dynamics} 
 
\author{Peter Mayer and Peter Sollich}

\affiliation{Department of Mathematics, King's College, Strand, 
London, WC2R 2LS, UK}
 
\date{\today} 
 
\begin{abstract} 
We present examples demonstrating that quasi-equilibrium
fluctuation-dissipation behavior at short time differences is not a 
generic feature of systems with slow non-equilibrium dynamics. We 
analyze in detail the non-equilibrium  fluctuation-dissipation 
ratio $X(t,\tw)$ associated with a defect-pair observable in the 
Glauber-Ising spin chain. It turns out that $X \neq 1$ throughout the 
short-time regime and in particular $X(\tw,\tw) = 3/4$ for 
$\tw\to\infty$. 
The analysis is extended to observables detecting defects at a finite
distance from each other, where similar violations of
quasi-equilibrium behaviour are found.
We discuss our results in the context of metastable 
states, which suggests that a violation of short-time quasi-equilibrium 
behavior could occur in general glassy systems for appropriately chosen 
observables. 
\end{abstract} 
 
 
\maketitle

\section*{INTRODUCTION}
\label{intro}

It is a common notion that in the non-equilibrium dynamics of glassy
systems, fluctuations of microscopic quantities are essentially at
equilibrium at short times, before the system displays aging. The 
concept of inherent structures or metastable states~\cite{StiWeb84}, 
for instance, is closely related to this picture. There, one partitions 
phase space into basins of attraction, e.g., corresponding to energy 
minima. At short times the system is expected to remain trapped in such 
metastable states, leading to equilibrium-type fluctuations, but activated 
inter-basin transitions eventually produce a signature of the underlying 
slow non-equilibrium evolution. 

A now standard procedure for  characterizing how far a system is from
equilibrium is provided 
by the non-equilibrium violation of the fluctuation-dissipation
theorem (FDT)~\cite{CugKur93,CugKurPel97,CriRit03}. Starting 
from a non-equilibrium initial state prepared at time $t=0$ one 
considers, for $t \geq \tw \geq 0$, 
\begin{equation}
  R(t,\tw) = X(t,\tw) \frac{\partial}{\partial \tw} C(t,\tw), 
  \label{equ:FDR}
\end{equation}
where $C(t,\tw) = \langle A(t) B(\tw) \rangle - \langle A(t) \rangle 
\langle B(\tw) \rangle$ is the connected two-time correlation 
function between some observables $A,B$ and $R(t,\tw) = T \left. 
\delta\langle A(t)\rangle/\delta h(\tw) \right|_{h=0}$ is the conjugate 
response function; the perturbation associated with the field $h$ is 
$\delta \mathcal{H} = - h(t) B$. Note that we have absorbed the 
temperature $T$ into our definition of the response function. Thus, for 
equilibrium dynamics, FDT implies that the fluctuation-dissipation ratio 
(FDR) $X(t,\tw)$ defined by (\ref{equ:FDR}) always equals one. 
Correspondingly, a parametric fluctuation-dissipation (FD) plot of the
susceptibility $\chi(t,\tw)=\int_{\tw}^t \dd\tau \, R(t,\tau)$ versus
$C(t,t)-C(t,\tw)$ has slope $X=1$.
In systems exhibiting slow dynamics and aging, on the other hand, the 
FDR $X(t,\tw)$ is often found to have a non-trivial scaling form 
in the limit of long times~\cite{CriRit03}. Over short observation intervals 
$\dt = t-\tw$ and at large waiting times $\tw$, however, one expects 
to recover equilibrium FDT with $X(t,\tw)=1$ according to, e.g., the 
inherent structure picture. Such ``quasi-equilibrium'' behavior at 
short times has been observed in essentially all previous 
studies~\cite{CriRit03} involving {\em microscopic} observables.  

Rigorous support is given to the quasi-equilibrium scenario 
in~\cite{CugDeaKur97}. There a general class of systems governed 
by dissipative Langevin dynamics is considered. Based on the 
entropy production rate in non-equilibrium dynamics, a bound 
is derived for the differential violation $V(t,\tw)$ of FDT, 
\begin{equation}
  V(t,\tw) = \frac{\partial C}{\partial \tw} - R = 
    \left[1-X(t,\tw)\right] \frac{\partial C(t,\tw)}{\partial\tw}. 
  \label{equ:V}
\end{equation}
The bound implies that under rather general assumptions~\cite{CugDeaKur97} 
and for observables only depending on a {\em finite} number of degrees 
of freedom one has $V(t,\tw) \to 0$ for any $\dt \geq 0$ fixed and 
$\tw\to\infty$; we call this the "short-time regime" from now on. 
Via~(\ref{equ:V}) it is concluded that $X(t,\tw) 
\to 1$ in the short-time regime. Clearly, however, the last 
step in this reasoning is only justified if 
$\lim_{\tw\to\infty} (\partial/\partial\tw) C(t,\tw)$ at fixed $\dt$
is non-vanishing. This will be the case if, for example, $C(t,\tw)$ 
admits a decomposition into stationary and aging parts~\cite{BouCugKurMez98} 
at large $\tw$, 
\begin{equation}
  C(t,\tw) = \Cst(t-\tw) + \Cag(t,\tw).
  \label{equ:Cstag}
\end{equation}
On the other hand, 
if $C(\tw,\tw)\to 0$ for $\tw\to\infty$ then also $\lim_{\tw\to\infty}
(\partial/\partial\tw) C(t,\tw)$ will normally vanish. For such
observables, $X(t,\tw)$ can {\em a priori} take arbitrary values 
in the short-time regime without violating the bound on 
$V(t,\tw)$ derived in~\cite{CugDeaKur97}. 

In (\ref{equ:Cstag}), the stationary contribution can be defined
formally as $\Cst(\dt)=\lim_{\tw\to\infty}C(\dt+\tw,\tw)$. If
$\Cst(\dt)$ has a nonzero limit for $\dt\to\infty$, then this is
conventionally included in $\Cag$ instead. One then
finds that the remainder, $\Cag$, is an ``aging'' function: for large
times it typically depends only on the ratio $t/\tw$.
Correlation functions of the form (\ref{equ:Cstag}) are often found in 
aging systems, for instance when considering local spin-observables 
$A=B=\sigma_i$ in critical ferromagnets~\cite{GodLuc02}; here $\Cag$ also
contains an overall $\tw$-dependent amplitude factor.
Further examples would be 
spin observables in $p$-spin models~\cite{CugKur93} or density fluctuations 
in MCT~\cite{GoetSjoe92}.
In either case the stationary correlations $\Cst(t-\tw)$ are intrinsically 
equilibrium-related [bulk fluctuations, dynamics within metastable states, 
"cage rattling"] and quasi-equilibrium behaviour is enforced by 
the bound of~\cite{CugDeaKur97}. Beyond that, however, there are
various examples
in the recent literature~\cite{MayBerGarSol03,BuhGar02b,Buhot03} 
displaying quasi-equilibrium behaviour even though 
$C(\tw,\tw) \to 0$ for $\tw \to \infty$. In 
the context of ferromagnets and spin-facilitated models such correlations 
are obtained when considering domain-wall or defect observables $A=B=n_i$. 
A generalization of (\ref{equ:Cstag}) accounting for the 
decrease of equal time correlations with $\tw$ is
\begin{equation}
  C(t,\tw) = A(\tw) \cst(t-\tw) + \Cag(t,\tw). 
  \label{equ:CAstag} 
\end{equation}
In the short-time contribution $\Cst(t,\tw) = A(\tw)
\cst(t-\tw)$~\footnote{We use the term ``short-time'' here rather
than ``stationary'' because of the $\tw$-dependent amplitude $A(\tw)$
in $\Cst$.}, $\cst$ is now defined as
$\cst(\dt)=\lim_{\tw\to\infty}C(\dt+\tw,\tw)/A(\tw)$. It is useful to fix
$\cst(0)=1$, so that $A(\tw)$ must behave asymptotically as
the equal-time correlation $C(\tw,\tw)$.
Results for defect observables in one and two-dimensional Ising 
models~\cite{MayBerGarSol03} are compatible with this scaling. 
For the one-dimensional FA and East models a similar picture emerges 
and it has been conjectured that aging contributions are in fact 
absent~\cite{BuhGar02b,Buhot03}. 
We discuss the scaling (\ref{equ:CAstag}) and its implications 
in more detail below. It is clear, however, that if $C(t,\tw) \sim 
A(\tw) \cst(t-\tw)$ and also $R(t,\tw) \sim A(\tw) \rst(t-\tw)$ 
in the short-time regime [where $\sim$ denotes asymptotic similarity 
for $\tw\to\infty$] then quasi-equilibrium behaviour 
requires~\footnote{
The term $[(\partial/\partial \tw)A(\tw)] \cst(t-\tw)$ in 
$(\partial/\partial \tw) \Cst(t,\tw)$ is subleading for amplitudes 
$A(\tw)$ with, e.g., power-law decay, and thus irrelevant when 
$\tw \to \infty$.} 
that $\rst(t-\tw) = (\partial/\partial \tw) \cst(t-\tw)$. The results 
of~\cite{MayBerGarSol03,BuhGar02b,Buhot03} indeed support this link. 

Recently, the authors of \cite{SasDorCha03} have exploited the notion
of quasi-equilibrium to {\em define} a nominal system temperature
$T_\mathrm{dyn}$ even for models where a thermodynamic bath
temperature $T$ does not a priori exist, e.g., because the dynamics
does not obey detailed balance. $T_\mathrm{dyn}$ is determined from the
short-time dynamics of correlations and responses, and the authors 
of~\cite{SasDorCha03} argue that for systems coupled to a heat bath 
this definition should generically reduce to $T_\mathrm{dyn}=T$. They 
in fact attempt a proof of this statement~\footnote{The argument for 
observable independence of $T_\mathrm{dyn}$ is 
based on observables of the form $O_n(\bsig) = \prod_{k=1}^n 
\sigma_{r_k}$. These have the features $O_n^2 = 1$ and $O_n(\bsig') = 
\pm O_n(\bsig)$ if $\bsig,\bsig'$ only differ by a single spin-flip. 
Using these identities the authors of~\protect\cite{SasDorCha03} show
that $T_\mathrm{dyn}$, derived from
two-time {\em auto}-correlation and response functions associated with 
$O_n$, is independent of the particular choice of $O_n$. They then 
claim that the same is true for linear combinations $\sum_{n\geq 1} 
a_n O_n$, which is wrong. Neither of the two features of $O_n$ just 
mentioned applies to linear combinations such that the proof breaks 
down. Also, disconnected instead of connected correlations are 
considered such that the whole argument only applies for observables 
$O_n$ with $n$ odd [where $\langle O_n \rangle = 0$ at all times].}, 
for spin models in the universality class of the two-dimensional Ising 
model. 

The dependence of $X(t,\tw)$ on the pair of observables $A,B$ used to
probe non-equilibrium FDT, Eq.~(\ref{equ:FDR}), in finite-dimensional
systems is still an actively debated 
issue~\cite{BerBar02,BerBar02b,FieSol02,SolFieMay02,MayBerGarSol03,CriRit03}. 
This is of particular
relevance as regards the possibility of characterizing the slow
dynamics in glassy systems by a time-scale dependent effective
temperature $T_\mathrm{eff}=T/X(t,\tw)$~\cite{CugKurPel97}. 
Beyond mean-field models one
does not expect $X(t,\tw)$, or more precisely its long time scaling,
to be completely robust against the choice of $A,B$. Instead, it has
been suggested that there may only be a limited class of ``neutral''
observables which allow a measurement of effective
temperatures~\cite{SolFieMay02,CriRit03}. It then seems plausible that
also the notion of quasi-equilibrium in the short-time regime may not
hold for all observables. This prompts us to revisit
the observable dependence of short-time fluctuation-dissipation
relations.

To be able to carry out explicit calculations, we study a simple
coarsening system, the one-dimensional ferromagnetic Ising spin chain
with Glauber dynamics. Coarsening systems are, of course, different
from glasses but they do exhibit aging behavior, easily interpretable
because of its link to a growing length scale. This makes them useful
``laboratory'' systems for testing general ideas and concepts
developed for systems with slow dynamics. The dynamical length scale
in a coarsening system -- which in our case is just the typical domain
size -- allows one to distinguish
equilibrated modes from slowly evolving non-equilibrium modes. For
{\em spatially localized} observables one thus naively expects
quasi-equilibrium dynamics at short times, as soon as the domain size
has become much 
larger than the length scale probed by the observable. 
But, as we will see in the following, this is 
not true in general: there are many local observables that do {\em not} 
obey the equilibrium FDT even in the short-time regime. 
Such nontrivial violations of quasi-equilibrium behaviour have 
to be distinguished from what is found for {\em global} 
observables such as, for example, the magnetization~\cite{MayBerGarSol03}: 
these depend on an extensive number of degrees of freedom, hence 
the bound of~\cite{CugDeaKur97} does not apply and one would not 
expect quasi-equilibrium behaviour at short times. In coarsening
systems this is also physically transparent: global observables
measure the dynamics on lengthscales larger than the
typical domain size, where equilibration has not yet taken place.

We analyze the non-equilibrium FDT in the Glauber-Ising chain
for observables that probe correlations between domain walls 
(defects) at distances $d \geq 1$. In Section~\ref{sec:prelim} we define 
our two-time correlation and response functions; their exact 
derivation for the case $d=1$ is sketched in the Appendix. 
We then recall some useful facts about the domain size distribution, 
both in and out-of equilibrium. In Sections~\ref{sec:eq} and~\ref{sec:noneq} 
we study adjacent defects, i.e.\ $d=1$. Some features
of the equilibrium dynamics, where FDT is obviously satisfied, are
discussed in Section~\ref{sec:eq}. The low temperature coarsening
dynamics are then analyzed in Section~\ref{sec:noneq} and compared to
the baseline provided by the equilibrium results. In particular, we focus 
on the short-time regime in
Sections~\ref{sec:short}--\ref{sec:FDplot}.
The aging behaviour is discussed in Section~\ref{sec:longtime} 
while Section~\ref{sec:equilibration} deals with the crossover to 
equilibrium. 
Based on the understanding developed for $d=1$, nonequilibrium 
FD relations for defects at distances $d>1$ are then studied in
Section~\ref{sec:multi}. We  
conclude in the final section with a summary and discussion.

\section{Defect pair observables}
\label{sec:prelim}

In order to obtain nontrivial fluctuation-dissipation behaviour in 
the short-time regime we have to consider non-standard observables; in the 
Glauber-Ising chain local spin as well as defect observables 
satisfy quasi-equilibrium~\cite{GodLuc00,LipZan00,MayBerGarSol03}. However, 
as already mentioned in~\cite{MayBerGarSol03}, multi-defect observables 
are potentially interesting candidates for new results. The simplest 
choice in this class are the defect-pair observables $A_d = n_i n_{i+d}$ 
with $d \geq 1$. We introduce the connected two-time autocorrelation 
functions associated with $A_d$ as  
\begin{eqnarray}
  C_d(t,\tw) &=& \langle n_i(t) n_{i+d}(t) n_i(\tw) n_{i+d}(\tw) \rangle 
  \nonumber \\
  & & - \langle n_i(t) n_{i+d}(t) \rangle 
    \langle n_i(\tw) n_{i+d}(\tw) \rangle. 
    \label{equ:cor2tdef}
\end{eqnarray}
The local two-time defect-pair response functions are 
\begin{equation}
  R_d(t,\tw) = T \left. \frac{\delta \langle n_i(t) n_{i+d}(t) \rangle}
  {\delta h(\tw)} \right|_{h=0}, 
  \label{equ:res2tdef}
\end{equation}
where the perturbation $\delta \mathcal{H} = - h(t) n_i n_{i+d}$ is applied. 
Throughout the paper we use the short-hands $C \equiv C_1$ and $R \equiv R_1$ 
for the case $d=1$. Our subsequent analysis of $C$ and $R$ is based on exact 
expressions, see Appendix. Since $A_d$ is nonzero only if we simultaneously 
have 
defects at sites $i$ and $i+d$, its behavior will reflect the domain size 
distribution in the system, and an understanding of the latter will be useful.

\subsection{Domain Size Distribution}
\label{sec:domain}

To summarize briefly, the Glauber-Ising chain~\cite{Glauber63} is defined 
on a one-dimensional lattice of Ising spins $\sigma_i = \pm 1$ with 
Hamiltonian $\mathcal{H}=-J \sum_i \sigma_i \sigma_{i+1}$, where 
each spin $\sigma_i$ flips with rate $w_i(\bsig)=\frac{1}{2}
[1-\frac{1}{2}\gamma
\sigma_i(\sigma_{i-1}+\sigma_{i+1})]$; here $\gamma=\tanh(2J/T)$. 
In terms of the domain-wall indicators or defect variables $n_i = \frac{1}{2} 
\left(1-\sigma_i \sigma_{i+1} \right)\in\{0,1\}$ the density of
domains $D_k$ of given size $k$ is expressed, using translational 
invariance, as
\begin{equation}
  D_k = \langle n_0 (1-n_1) \cdots (1-n_{k-1}) n_k \rangle. 
  \label{equ:Dkdef}
\end{equation}
As usual $\langle \, \cdot \, \rangle$ refers to the ensemble average
in the case of equilibrium and otherwise to an average over initial
configurations and stochasticity in the dynamics.

In equilibrium the derivation of $D_k$ for the Glauber-Ising chain is 
straightforward: from 
$\langle \prod_i n_i \rangle = \prod_i \langle n_i \rangle$ and  
translational invariance we have 
\begin{equation}
  D_{k,\eq} = \langle n_0 \rangle^2 \left[ 1- \langle n_0 \rangle
    \right]^{k-1}.
  \label{equ:Dkeq}
\end{equation}
The distribution of domain sizes $k$ is thus exponential in equilibrium, 
with the most frequent domains those of size one. 
The mean domain
size, on the other hand, is given by the inverse of the concentration
of defects $\langle n_0 \rangle$. One easily shows that 
\begin{equation}
  c_{\eq}(\taueq) = \langle n_0 \rangle = \frac{1}{\sqrt{2\taueq}} \, 
  \frac{\sqrt{1+\gamma}-\sqrt{1-\gamma}}{\sqrt{2} \, \gamma}. 
  \label{equ:ceq}
\end{equation}
For equilibrium quantities we generally
use the equilibration time $\taueq=1/(1-\gamma) \sim \frac{1}{2} \exp(4J/T)$ 
to parametrize
temperature.  At low $T$ the defect concentration scales as $c_{\eq}
\sim (2\taueq)^{-1/2}$. Hence the mean domain size is 
$O\left(\sqrt{\taueq}\right)$.
From (\ref{equ:Dkeq}), (\ref{equ:ceq}) 
the density of small domains with size $k=O(1)$ 
is then flat, $D_{k,\eq} \sim 1/(2\taueq)$.

We note briefly that our $D_k$ are densities of domains of given 
size, rather than a normalized domain size distribution. The normalization
factor is simply the defect concentration since, from
(\ref{equ:Dkeq}), $\sum_{k=1}^\infty D_k = \langle n_0\rangle$. 
Abbreviating $c=\langle n_0\rangle$, we can thus write $D_k=cP_k$ with 
$\sum_{k=1}^\infty P_k=1$. For small $c$, the normalized distribution $P_k$ 
often assumes a scaling form, with $k$ scaled by the mean domain size $1/c$:
$P_k = c P(\kappa)$ with $\kappa=kc$, and correspondingly $D_k = c^2
P(\kappa)$. From (\ref{equ:Dkeq}), the equilibrium scaling function
is exponential, $P(\kappa)=\exp(-\kappa)$.

For the out-of-equilibrium case a derivation of $D_k$ is rather less
trivial; a corresponding calculation for the $1d$ Potts model is 
given in \cite{DerZei96}.  For the Ising case and a quench at time
$t=0$  from a 
random, uncorrelated initial state to zero temperature, the
results are as follows: the mean domain size grows as
$c^{-1}=O\left(\sqrt{t}\right)$ with typical domains having a
concentration $O(c^2)=O\left(t^{-1}\right)$. For large domain sizes,
$k \gg \sqrt{t}$, $D_k(t)$ has an exponential tail $O\left((1/t)
\exp(-\alpha \, k/\sqrt{t})\right)$; an expression for the constant $\alpha$
is given in \cite{DerZei96}. The density of small domains $k \ll
\sqrt{t}$, on the other hand, is linearly related to the domain size,
with $D_k(t) = O\left( k/t^{3/2}\right)$. Correspondingly, the scaling
form $P(\kappa)$ of the normalized domain size distribution decays
exponentially for large $\kappa$, but is linear in $\kappa$ for
$\kappa\ll 1$. 

The precise scaling of the density of small domains $k=O(1)$ in nonequilibrium 
coarsening is easily derived. Instead of directly working out the $D_k$, 
which is cumbersome, consider for a moment the quantity 
\begin{equation}
  \widetilde{D}_k = \langle n_0 (1-2n_1) \cdots (1-2n_{k-1}) n_k \rangle. 
  \label{equ:Pkdef}
\end{equation}
In contrast to the $D_k$, any $\widetilde{D}_k$ can be conveniently expanded 
in terms of two-spin correlations. We have, in fact, $\widetilde{D}_k = 
\frac{1}{4} \langle \sigma_1 \sigma_k-\sigma_0 \sigma_k-\sigma_1 
\sigma_{k+1}+\sigma_0 \sigma_{k+1} \rangle$. 
For zero-temperature coarsening one shows~\cite{MaySol04}
\begin{equation}
  \widetilde{D}_k(t) = \frac{k}{2} \, \me^{-2t}\frac{I_k(2t)}{2t},
  \label{equ:P}
\end{equation}
where the $I_n(x)$ are modified Bessel functions~\cite{Mathbook}. 
Now compare the definitions of $D_k$ and $\widetilde{D}_k$ in the limit of 
large 
times. For both quantities we have that only states with $n_0 = n_k = 1$ 
contribute. To leading order these states do not contain any further 
defects $n_i$ in the range $i = 1,\ldots, k-1$, hence 
$D_k \sim \widetilde{D}_k \sim k/\left(8 \sqrt{\pi} t^{3/2} \right)$ for 
$t \to \infty$. States that do contain further defects in this range, 
on the other hand, cause $D_k$ to differ from $\widetilde{D}_k$. In an 
independent interval approximation, which gives the correct scaling but 
incorrect prefactors~\cite{AleBen95,DerZei96}, the chances to have an 
additional defect 
at site $i$ are $D_i D_{k-i} = O\left(t^{-3}\right)$. Contributions from 
states containing more than one defect in this range are even smaller,
giving overall $D_k = \widetilde{D}_k + O\left(t^{-3}\right)$. By the same 
reasoning we also 
have $\langle n_i n_{i+k} \rangle = \widetilde{D}_k + O\left(t^{-3}\right)$. 
These 
scalings apply for any fixed $k \geq 2$ and in the limit of large $t$. 
For $k=1$, finally, we have $\langle n_i n_{i+1} \rangle = D_1 = 
\widetilde{D}_1$ 
as the definitions coincide in this case. 

Note that when comparing only the scale of typical
domains in and out of equilibrium, an out-of-equilibrium system of age
$t$ is comparable to an equilibrium system with equilibration
time $\taueq\approx t$. Indeed, typical domains have
size $O\left(\sqrt{\taueq}\right)$ and density
$O\left(\taueq^{-1}\right)$ in equilibrium, while out
of equilibrium the same quantities scale like 
$O\left(\sqrt{t}\right)$ and $O\left(t^{-1}\right)$, respectively.
However, this correspondence does not extend to the
details of the shape of the domain size distribution. In particular,
it breaks down for small domains $k=O(1)$. In equilibrium such domains
have a concentration $O\left(\taueq^{-1}\right)$ while in the
corresponding coarsening situation their concentration
$O\left( k/t^{3/2}\right)$ is much smaller. 

It is instructive to note that Glauber 
dynamics for the spin system $\{\sigma_i\}$ corresponds to a diffusion limited 
reaction process~\cite{Santos97} for the defects $\{n_i\}$; the diffusion rate 
is $\frac{1}{2}$. 
At low $T$ adjacent defects annihilate with rate close to one while pair 
creation, i.e., flipping a spin within a domain, occurs with rate 
$1/(2\taueq)$. The latter process is important in equilibrium -- continuously 
producing new domains of size one -- but is unimportant
at low temperatures while the system is coarsening, and indeed strictly
absent at zero temperature. This leads to the 
different scalings of the 
density of small domains in and out of equilibrium.

\section{Equilibrium}
\label{sec:eq}

In order to familiarize ourselves with the dynamics of defect pairs 
we now study the equilibrium behavior of the two-time 
correlation $C(t,\tw)$. An exact expression is obtained from 
the result for a quench to finite temperature given in 
the Appendix by taking the limit $\tw \to \infty$ 
at fixed $\dt = t-\tw$. We use the notation 
$C_{\eq}(\dt,\taueq) = \lim_{\tw\to\infty} C(\dt+\tw,\tw)$ for 
the equilibrium correlation; from (\ref{equ:cor2t}) one has 
\begin{eqnarray}
  C_{\eq}(\dt,\taueq) & = & \frac{1}{2\gamma\taueq} \Hn_{1,\eq}(\dt) 
    \bigg\{ \me^{-\dt} [ \In_0 - \In_1](\gamma \dt) 
    \nonumber \\ 
  & & - \frac{1}{2\gamma\taueq} \Hn_{1,\eq}(\dt) \bigg\}. 
    \label{equ:coreq}
\end{eqnarray}
Here we have introduced the short hand $[ \, \cdot \, ](x)$ to indicate 
that all functions enclosed in the square brackets have argument $x$. 
The function $H$ is introduced in (\ref{equ:Hn}) and discussed in the 
Appendix. 
Because FDT is satisfied in equilibrium the conjugate response to 
(\ref{equ:coreq}) 
is $R_{\eq}(\dt,\taueq) = - \partial_{\dt} C_{\eq}(\dt,\taueq)$.
Consequently the equilibrium susceptibility is given by 
\begin{equation}
  \chi_{\eq}(\dt,\taueq) = C_{\eq}(0,\taueq) - C_{\eq}(\dt,\taueq), 
  \label{equ:chieq}
\end{equation}
and we subsequently focus on the discussion of $C_{\eq}$.

\subsection{Small $\dt$ Regime}
\label{sec:eqst}

Let us first consider the dynamics of $C_{\eq}(\dt,\taueq)$ for finite 
$\dt \geq 0$ and in the limit of low temperatures $\taueq \gg 1$. 
Via the definition (\ref{equ:cor2tdef}) of $C \equiv C_1$ the equal-time value 
$C_{\eq}(0,\taueq) = \langle n_i n_{i+1} \rangle - \langle n_i n_{i+1} 
\rangle^2$ is directly linked to the density of domains of size one, 
$D_{1,\eq}(\taueq) = \langle n_i n_{i+1} \rangle$. From (\ref{equ:Dkeq}), 
(\ref{equ:ceq}) and setting $\dt=0$ in (\ref{equ:coreq}) 
\begin{equation}
  D_{1,\eq}(\taueq) 
  = \frac{1}{2\taueq} \frac{1-\sqrt{1-\gamma^2}}{\gamma^2}
  = \frac{1}{2\gamma\taueq} \Hn_{1,\eq}(0).  
  \label{equ:D1eq}
\end{equation}
At low temperatures $D_{1,\eq}(\taueq)$ and thus 
$C_{\eq}(0,\taueq)$ scales as $D_{1,\eq}(\taueq) \sim 1/(2\taueq)$. 
Now, for finite $\dt > 0$ and in the limit of low temperatures 
$\taueq\gg 1$ an expansion of (\ref{equ:coreq}) gives, to leading 
order, 
\begin{equation}
  C_{\eq}(\dt,\taueq) = p_1(\dt) D_{1,\eq}(\taueq)  + 
  O\left(\taueq^{-2}\right), 
  \label{equ:coreqdtO1}
\end{equation}
where
\begin{equation}
  p_1(\dt) = \me^{-2\dt} [\In_0^2 - \In_1^2](\dt).
  \label{equ:prD}
\end{equation}
From (\ref{equ:coreqdtO1}) and our knowledge of the equilibrium
domain size distribution we may assign a direct physical meaning to
$p_1(\dt)$: in the limit of low temperatures and at finite $\dt$ the
connected and disconnected correlations coincide to leading order,
i.e., $C_{\eq}(\dt,\taueq) \sim \langle n_i(\dt) n_{i+1}(\dt) n_i(0)
n_{i+1}(0) \rangle$. So only situations where sites $i$ and $i+1$ are
occupied by defects at both times contribute to $C_{\eq}$. But since
the size of typical domains scales as $O\left(\sqrt{\taueq}\right)$,
the probability for neighboring domains to be of size $O(1)$ vanishes
at low temperatures. Therefore, and since $\dt = O(1)$, the defect
pair at sites $i,i+1$ at the later time $\dt$ must in fact be the one
that also occupied these sites at the reference time.  Hence we may
interpret $p_1(\dt)$ as the ``random walk return probability of an
adjacent defect pair". 

This scenario is easily verified by direct 
calculation. Consider a one-dimensional lattice containing exactly 
two defects at sites $k$ and $l$ with $k < l$ at time $t = 0$. Denote 
by $p_t(i,j)$ the probability to find these defects at sites $i < j$ 
at time $t$. Since the $T=0$ dynamics of defects in the Glauber-Ising 
chain is diffusion-limited pair annihilation~\cite{Santos97} with 
diffusion rate 
$\frac{1}{2}$ and annihilation rate one the $p_t(i,j)$ 
satisfy~\cite{DerZei96} 
\begin{eqnarray}
  \frac{\partial}{\partial t} p_t(i,j) & = & -2 p_t(i,j) + \frac{1}{2} 
  \left[ p_t(i-1,j) + p_t(i+1,j) \right. \nonumber \\ 
  & & \left. + p_t(i,j-1) + p_t(i,j+1) \right].
\end{eqnarray}
This system of equations must be solved over $i < j$ subject to the 
boundary conditions $p_t(i,i)=0$. Using images~\cite{DerZei96, MaySol04} 
it is straightforward to show that the solution is 
\begin{equation}
  p_t(i,j) = \sum_{k < l} G_t(i,j;k,l) p_0(k,l), 
\end{equation}
where 
\begin{equation}
  G_t(i,j;k,l) = \me^{-2t} \left[ I_{i-k} I_{j-l} - I_{i-l} I_{j-k} 
  \right](t). 
  \label{equ:G}
\end{equation}
It is clear from this result that 
the Green's function $G_t(i,j;k,l)$ is in fact the conditional probability 
of finding the defect-pair at sites $i < j$ at time $t$ given that it
was initially
located at sites $k < l$. Consequently $p_1(t) = G_t(i,i+1;i,i+1)$ as claimed 
above. The two-time defect-pair correlation $C_{\eq}(\dt,\taueq)$, 
Eq.~(\ref{equ:coreqdtO1}), is thus to leading order given by the probability 
$D_{1,\eq}(\taueq)$ of having a defect pair at sites $i,i+1$ times the 
conditional 
probability $p_1(\dt)$ for this pair also to occupy the same sites a
time $\dt$
later. For $\dt \gg 1$ an expansion of (\ref{equ:prD}) gives 
$p_1(\dt) \sim 1/(2\pi\dt^2)$: the return probability for the defect 
pair drops quite rapidly as defects are likely to have disappeared via 
annihilation in the time interval $[0,\dt]$ if $\dt \gg 1$.

\subsection{Large Times}
\label{sec:eqlargetimes}

When $\dt$ becomes comparable to $\taueq$ the simple picture discussed above 
breaks down; annihilation events with remote defects and pair creation
are then relevant. But from equation (\ref{equ:coreq}) 
results for this regime, which are formally obtained by taking 
$\dt,\taueq \to \infty$ with their ratio fixed, 
are easily derived. In this limit we replace the modified Bessel functions 
appearing in (\ref{equ:coreq}) with their asymptotic 
expansions~\cite{Mathbook}. 
This produces the leading order scalings 
\begin{eqnarray}
  C_{\eq}(\dt,\taueq) & \sim & \frac{1}{4\pi\dt^2\taueq} \qquad \qquad
  [\dt \ll \taueq], 
  \label{equ:coreqdtlltaueq}
  \\ 
  C_{\eq}(\dt,\taueq) & \sim & \frac{3\taueq}{16\pi\dt^4} \me^{-2\dt/\taueq}
  \hspace{1.84ex} [\dt \gg \taueq]. 
  \label{equ:coreqdtggtaueq}
\end{eqnarray}
The expansion (\ref{equ:coreqdtlltaueq}) matches the large
$\dt$ limit of (\ref{equ:coreqdtO1}). So up to the time scale $\dt =
O(\taueq)$ the decay of the {\em connected} two-time defect pair
correlation is controlled by the defect pair return probability. 
For times $\dt$ beyond $O(\taueq)$, defect configurations
are reshuffled via pair creation and the connected correlation 
vanishes exponentially as one might expect. For later reference 
we note that according to (\ref{equ:coreqdtO1}), (\ref{equ:coreqdtlltaueq}),
(\ref{equ:coreqdtggtaueq}) we have $C_{\eq}(\dt,\taueq) > 0$ at all
times.

There is, however, a subtle effect in 
the underlying physics. This becomes obvious when considering 
disconnected correlation functions.  The disconnected defect pair
correlation in equilibrium is $C_{\eq}^{\mathrm{DC}}(\dt,\taueq) =
\langle n_i(\dt) n_{i+1}(\dt) n_j(0) n_{j+1}(0) \rangle$, and is
linked to the connected one via $C_{\eq}^{\mathrm{DC}}(\dt,\taueq) =
C_{\eq}(\dt,\taueq) + D_{1,\eq}^2(\taueq)$. Now according to
(\ref{equ:D1eq}) we have $D_{1,\eq}^2(\taueq) =
O\left(\taueq^{-2}\right)$ while from (\ref{equ:coreqdtlltaueq}),
$C_{\eq}(\dt,\taueq) = O\left(\dt^{-2}\taueq^{-1}\right)$ for
$\dt\ll\taueq$. Therefore, if $\dt \gg \sqrt{\taueq}$,
$C_{\eq}(\dt,\taueq)$ is negligible compared to $D_{1,\eq}^2(\taueq)$
and so the disconnected correlation
$C_{\eq}^{\mathrm{DC}}(\dt,\taueq)$ becomes $\dt$-independent. In
other words, because of the rapid decay of the defect pair return
probability $p_1(\dt)$ we are more likely to find an independent
defect pair at sites $i$, $i+1$, rather than the original one, already
on a time scale $\dt = O\left(\sqrt{\taueq}\right)$. This is in marked
contrast to spin or (single) defect observables~\cite{MaySol04}, where 
this crossover happens on the time scale $\dt = O\left(\taueq\right)$.

Let us finally consider the equilibrium defect pair susceptibility 
$\chi_{\eq}(\dt,\taueq)$. According to (\ref{equ:chieq}) it is strictly 
increasing, implying that $R_{\eq}(\dt,\taueq) > 0$ at all times, 
and grows from its initial value of zero at $\dt=0$ to the asymptotic 
value $C_{\eq}(0,\taueq)$ on an $O(1)$ time scale. 
Explicitly we have from (\ref{equ:coreqdtO1}) the
approximation $\chi_{\eq}(\dt,\taueq) \approx 
D_{1,\eq}(\taueq) [1-p_1(\dt)]$ which holds uniformly in $\dt$ at low
temperatures.

\section{Non-Equilibrium}
\label{sec:noneq}

In this section we analyze defect pair correlation and response functions 
for zero temperature  coarsening dynamics following a quench from a random, 
uncorrelated initial state. For the most part we will focus on the short-time 
behavior of these functions. Following our discussion in the 
introduction we decompose the two-time functions into short-time and aging 
contributions, 
\begin{eqnarray}
  C(t,\tw) & = & \Cst(t,\tw) + \Cag(t,\tw), \label{equ:Csa} \\
  R(t,\tw) & = & \Rst(t,\tw) + \Rag(t,\tw). \label{equ:Rsa} 
\end{eqnarray}
The two-time correlation and response functions are obtained 
from (\ref{equ:cor2t}) and the construction of the response given in 
the Appendix. For mathematical simplicity we take the limit $T\to 0$
but stress that the results are also valid for nonzero temperatures $T
> 0$ while the system is still far from equilibrium; as discussed in
Sec.~\ref{sec:equilibration}, this requires $\tw,t\ll {\taueq}^{2/3}$.
The response functions are always derived by
taking the perturbing field $h$ to zero {\em before} taking $T\to
0$. This has to be done to ensure linearity of the response: as
discussed in more detail below, the size of the linear regime in the field
strength $h$ scales as $T$ for low $T$.
We find 
\begin{eqnarray}
  \Cst(t,\tw) & = & 
 	  \frac{1}{2} \me^{-2t} [\In_0^2 - \In_1^2](t-\tw) 
          \frac{\In_1(2\tw)}{2\tw}, 
 	  \label{equ:Cst} \\
  \Rst(t,\tw) & = & 
    \frac{1}{4} \me^{-2t} [(\In_0^2-\In_1^2) - \In_1 (\In_0-\In_2)](t-\tw) 
    \nonumber \\
    & & \times \frac{[\In_1+2\In_2](2\tw)}{2\tw}, 
    \label{equ:Rst}
\end{eqnarray}
\begin{widetext}
\begin{eqnarray}
  \Cag(t,\tw) & = & 
    - \frac{1}{4} \me^{-2(t+\tw)} \frac{\In_1^2(t+\tw)}{(t+\tw)^2} \nonumber \\
    & & + 
    \frac{1}{2} \me^{-2(t+\tw)} [\In_0-\In_1](t-\tw)
    \left\{ 
    \frac{\In_1(t+\tw)}{t+\tw} [\In_0+\In_1](2\tw)-[\In_0+\In_1](t+\tw) 
    \frac{\In_1(2\tw)}{2\tw} \right\}, 
    \label{equ:Cag} \\
  \Rag(t,\tw) & = & 
    \frac{1}{8} \me^{-2(t+\tw)} [\In_0-2\In_1+\In_2](t-\tw) 
    \left\{ \frac{[\In_1+2\In_2](t+\tw)}{t+\tw} [\In_0+\In_1](2\tw) - 
    [\In_0+\In_1](t+\tw) \frac{[\In_1+2\In_2](2\tw)}{2\tw} \right\}
    \nonumber \\
    & & + \frac{1}{4} \me^{-2(t+\tw)} [\In_1-\In_2](t-\tw) 
    \left\{ 
    \frac{\In_1(2\tw)}{2\tw} \frac{\In_1(t+\tw)}{t+\tw} - 
    \frac{[\In_1-\In_2](t+\tw)}{t+\tw} [\In_0+\In_1](2\tw) \right\}.
    \label{equ:Rag}
\end{eqnarray}
\end{widetext}
\begin{figure*}
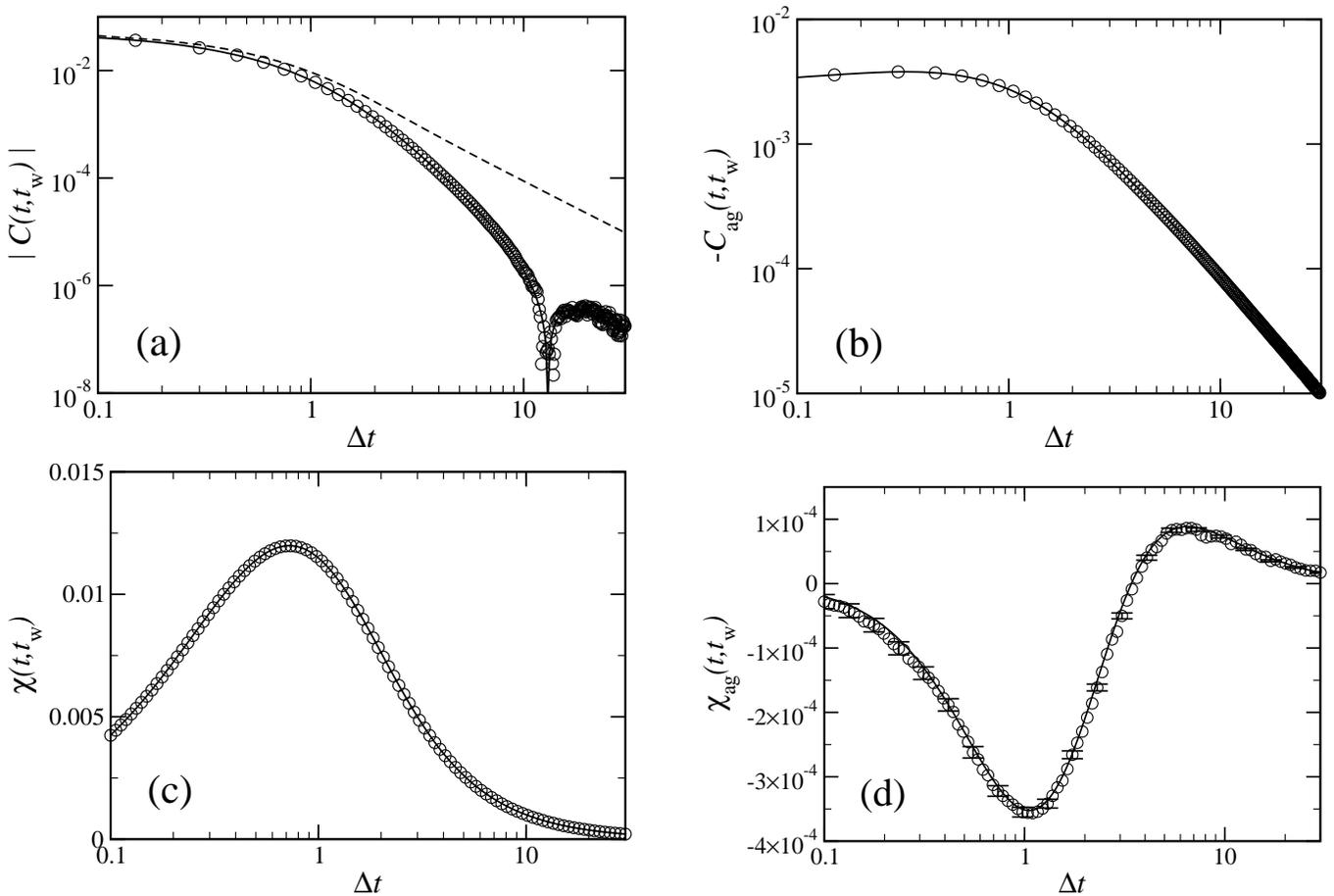

  \epsfig{file=ctotal.eps, width=8.5cm, clip} \hfill 
  \epsfig{file=caging.eps, width=8.5cm, clip} \\[1ex] 
  \epsfig{file=stotal.eps, width=8.5cm, clip} \hfill 
  \epsfig{file=saging.eps, width=8.5cm, clip}
  \caption{\label{fig:dat_sim} Comparison of our exact predictions 
    (\ref{equ:Csa}-
    \ref{equ:Rag}), shown as full curves, and simulation data [circles] 
    at $T=0$ 
    and $\tw = 1$. Predictions for the susceptibilities are obtained
    by numerical integration of  
    (\ref{equ:Rst}) and (\ref{equ:Rag}). The data for aging contributions in 
    panel (b) and (d) is obtained from the simulation data of $C(t,\tw)$ and 
    $\chi(t,\tw)$ shown in (a), (c) by subtracting off the exact short-time 
    contributions (\ref{equ:Cst}) [dashed line in panel (a)] 
    and the susceptibility corresponding to 
    (\ref{equ:Rst}). In panel (b) and (c) error bars for the simulation data 
    are invisible on the scale of the plot. The same is true for (a) 
    and $\dt < 
    10$. For $\dt > 10$, on the other hand, the standard deviation of the data 
    is around $10^{-7}$, which is reflected in the plot. In panel (d), 
    finally, 
    error bars [$\pm$ standard deviation] are shown for every fifth
    data point but are no larger than the symbols. 
   }
\end{figure*}
As we will see in Section~\ref{sec:short} below, only 
the short-time functions (\ref{equ:Cst}), (\ref{equ:Rst}) contain terms 
that contribute to leading order in the short-time regime. 

The results given above are exact. Before proceeding, we nevertheless
compare them with simulation data to exclude the possibility of trivial 
algebraic errors 
in the derivation and confirm some of the more surprising features that are 
discussed below.
For measuring two-time susceptibilities $\chi_{A,B}(t,\tw)$, with $A_i,B_i$ 
generic local observables, we use the standard 
method~\cite{Barrat98} of perturbing the system with $\delta\mathcal{H} = -h 
\sum_i \epsilon_i B_i$ for $t \geq \tw$ such that $\chi_{A,B}(t,\tw) = (T/h) 
\sum_i 
\left[\epsilon_i \langle A_i(t) \rangle \right]_\epsilon$ in the limit 
$h \to 0$. 
Here the $\epsilon_i \in \{-1,1\}$ are 
independent, identically distributed random variables and 
$[\,\cdot\,]_\epsilon$ 
denotes an average over their distribution.
Note that as our definition 
of the response contains a factor of $T$ this is also the case for 
$\chi_{A,B}(t,\tw)$. Transition rates in the presence of the perturbation are, 
to linear order in $h/T$, 
\begin{eqnarray}
  w^{(h)}_n(\bsig) & = & w_n(\bsig) + \frac{h}{T} w_n(\bsig) \left[ 
  1-w_n(\bsig) \right] \sum_m \epsilon_{n-m} \nonumber \\
  & & \times  \left[ B_{n-m}(F_n \bsig) - B_{n-m}(\bsig) \right].
  \label{equ:wh}
\end{eqnarray}
Here $F_n \bsig = (\ldots, -\sigma_n, \ldots)$ denotes the spin-flip operator 
and $w_n(\bsig)$ are standard Glauber rates without perturbation. 
In the procedure of measuring $\chi_{A,B}(t,\tw)$ the perturbing field $h$ 
only appears in the combination $\bar{h} = h/T$. There is therefore a 
well defined zero-temperature limit at fixed $\bar{h}$; the linear 
susceptibility is then obtained by using a sufficiently small
$\bar{h}$. We note finally that local perturbations only  
produce a few non-zero terms in the sum in~(\ref{equ:wh}). The 
defect pair observables $A_i = B_i = n_i n_{i+1}$ we are considering, 
for instance, only depend on $\sigma_i,\sigma_{i+1},\sigma_{i+2}$ and 
thus only $m=0,1,2$ contribute to the sum in~(\ref{equ:wh}).

We show in Section~\ref{sec:longtime} 
that with increasing $\tw$ it becomes harder to see 
aging contributions in the two-time correlation and response functions. 
In order to be able to resolve these aging contributions in a simulation 
we have chosen $\tw = 1$ and $\dt = 0.1 \ldots 30$. The data presented 
in Fig.~\ref{fig:dat_sim} are for zero-temperature simulations in a system 
of $10^6$ spins, averaged over $10^4$ runs. For measuring the 
susceptibility we use $\bar{h} = 0.2$, which is well within the linear
regime. Our code uses an event driven 
algorithm~\cite{BorKalLeb75} although for the small times considered here a 
standard 
Monte-Carlo method would be just as efficient. 
As a full discussion of $C(t,\tw)$ and $R(t,\tw)$ 
will be given in the subsequent sections we comment only briefly 
on the data in Fig.~\ref{fig:dat_sim}. According to our decomposition 
(\ref{equ:Csa}) we may think of the connected two-time defect pair correlation 
in Fig.~\ref{fig:dat_sim}(a) as the
sum of short-time and aging contributions.
From (\ref{equ:Cst}) short-time contributions are always positive while 
the aging contributions (\ref{equ:Cag}) are in fact negative, 
see Fig.~\ref{fig:dat_sim}(b). From Fig.~\ref{fig:dat_sim}(a) the 
correlation $C(t,\tw)$ is dominated by short-time contributions up to about 
$\dt = 1$. In the range $\dt = 1\ldots 13$ short-time and aging contributions 
compete, leading to a fast drop in $C(t,\tw)$. At $\dt \approx 13$,
$C(t,\tw)$ crosses zero, giving the cusp in the plot.
For 
$\dt > 13$ the two-time correlation is negative, with short-time and 
aging contributions almost cancelling each other. The two-time defect pair 
susceptibility $\chi(t,\tw)$ shown in Fig.~\ref{fig:dat_sim}(c),
may similarly be regarded as containing short-time and aging contributions 
according to~(\ref{equ:Rsa}). From Fig.~\ref{fig:dat_sim}(d) aging 
contributions 
in $\chi(t,\tw)$ are tiny such that a plot of $\Sst(t,\tw)$ alone would fit 
the data shown in Fig.~\ref{fig:dat_sim}(c) rather well. The small deviations 
of $\chi(t,\tw)$ from $\Sst(t,\tw)$, that is $\Sag(t,\tw)$, however, 
are precisely 
as predicted by (\ref{equ:Rag}), see Fig.~\ref{fig:dat_sim}(d). Altogether, 
the 
data presented in Fig.~\ref{fig:dat_sim} are fully consistent with,
and thus confirm, our 
exact results (\ref{equ:Cst}-\ref{equ:Rag}). We therefore now 
turn to a discussion of $C(t,\tw)$, $\chi(t,\tw)$ based directly on 
(\ref{equ:Cst}-\ref{equ:Rag}).

\subsection{Short-Time Regime}
\label{sec:short}

Here we analyze the dynamics of $C(t,\tw)$ and $R(t,\tw)$ in the 
short-time regime of $\dt \geq 0$ fixed and $\tw \to \infty$. For the 
correlation we have from an expansion of (\ref{equ:Cag}) that the aging
contribution scales as $\Cag(t,\tw) = O\left(\tw^{-3}\right)$ in 
this limit; already at $\tw = 10$ a plot of $\Cag$ would lie below
the vertical range of Fig.\ref{fig:dat_sim}(b). The term $\Cst(t,\tw)$,
Eq.~(\ref{equ:Cst}),  
on the other hand, is simply $\Cst(t,\tw) = p_1(t-\tw) D_1(\tw)$ as follows 
from $D_1 = \widetilde{D}_1$ and (\ref{equ:P}), (\ref{equ:prD}). 
Because $D_1(\tw) = 
O\left({\tw}^{-3/2}\right)$, aging contributions in $C(t,\tw)$ are 
subdominant. 
In the short-time regime the connected two-time defect pair correlation 
is thus to leading order given by $\Cst(t,\tw)$ alone, 
\begin{equation}
  C(t,\tw) = p_1(t-\tw) D_1(\tw) + O\left( \tw^{-3} \right). 
  \label{equ:coragdtO1}
\end{equation}
This scaling property is the key that separates short-time contributions 
$\Cst$ from aging terms $\Cag$ in (\ref{equ:Csa}), (\ref{equ:Cst}), 
(\ref{equ:Cag}). Furthermore $\Cst$ is of the form $\Cst(t,\tw) = 
A(\tw) \cst(t-\tw)$ as proposed in (\ref{equ:CAstag}). Comparison of 
(\ref{equ:coreqdtO1}) and (\ref{equ:coragdtO1}) shows that the nonequilibrium 
relaxation function $\cst(\dt) = p_1(\dt)$ {\em coincides} with its 
low-temperature equilibrium counterpart. Only the amplitude $A$ 
is given by the dynamical density of defect pairs $D_1(\tw)$ instead 
of its equilibrium analogue $D_{1,\eq}(\taueq)$. 

The result (\ref{equ:coragdtO1}) 
is easily explained by random walk arguments, in full analogy to 
Sec.~\ref{sec:eqst}. Connected and disconnected correlations coincide 
to $O\left(D_1^2\right) = O\left(\tw^{-3}\right)$. For the disconnected 
correlation to be nonzero we need states containing a defect pair at sites 
$i,i+1$ at time $\tw$. These occur with probability $D_1(\tw)$ and to order 
$O\left(\tw^{-3}\right)$ there are no further defects in any finite 
neighbourhood. Hence only if these defects also occupy the same sites 
at time $t$ is there a contribution to the disconnected correlation. This
occurs with probability $p_1(\dt) = G_{\dt}(0,1;0,1)$ and 
(\ref{equ:coragdtO1}) follows. 

Now we turn to the behaviour of the defect pair response in the 
short-time regime. Expanding (\ref{equ:Rag}) shows that $\Rag = 
O\left( \tw^{-3} \right)$. But $\Rst = O\left( {\tw}^{-3/2} \right)$ 
from (\ref{equ:Rst}) so $R$ is dominated by $\Rst$ in the  
short-time regime, i.e., $R(t,\tw) = \Rst(t,\tw) + O\left( \tw^{-3} 
\right)$ just as was the case for the correlations. We rearrange 
this result into the form 
\begin{eqnarray}
  R(t,\tw) & = & \frac{3}{4} \left[ \frac{\partial}{\partial \tw} 
  p_1(t-\tw) -2 \me^{-2(t-\tw)} \frac{I_1^2(t-\tw)}{t-\tw} \right] 
  \nonumber \\ 
  & & \times  D_1(\tw) + O\left( \tw^{-5/2} \right),  
  \label{equ:resagdtO1}
\end{eqnarray}
where the expression in the square bracket coincides with the
$\dt$-dependent factor in (\ref{equ:Rst}). The $\tw$-dependent amplitude 
factor in (\ref{equ:Rst}) was expressed in terms of $D_1 = \widetilde{D}_1$, 
Eq.~(\ref{equ:P}), using $\me^{-x}\In_2(x) = \me^{-x}\In_1(x) + 
O\left(x^{-3/2}\right)$. Writing $\Rst(t,\tw) = A(\tw) \rst(t-\tw)$ 
as the equivalent of (\ref{equ:CAstag})
then shows that the nonequilibrium function $\rst(\dt)$ is 
{\em different} from its equilibrium counter part $(-\partial/\partial \dt) 
\cst(\dt) = (-\partial/\partial \dt) p_1(\dt)$. In fact, from 
$(\partial/\partial \tw) 
C(t,\tw) \sim [(\partial/\partial \tw) p_1(t-\tw)] D_1(\tw)$ and 
(\ref{equ:resagdtO1}) we obtain for the FDR in the short-time regime, 
abbreviating $X(\dt) = \lim_{\tw \to \infty} X(t,\tw)$, 
\begin{equation}
  X(\dt) = \frac{3}{4} \left[ 1 - \frac{2}{(-\partial/\partial \dt) p_1(\dt)}
  \me^{-2\dt} \frac{I_1^2(\dt)}{\dt} \right].
  \label{equ:Xqe}
\end{equation}
Thus the FDR is neither equal to one nor even constant in the 
short-time regime. In particular, for $\dt \to 0$ one finds 
\begin{equation}
  X(0) = \lim_{\tw \to \infty} X(\tw,\tw) = \frac{3}{4}.
  \label{equ:X34}
\end{equation}

\subsection{The Response Function}
\label{sec:response}

Let us now try to understand the origin for the anomalous short-time response 
(\ref{equ:resagdtO1}). For correlations we saw that the asymptotic
equality between $C$ and $\Cst$, 
Eq.~(\ref{equ:coragdtO1}), could be easily explained by random walk arguments. 
This is, in fact, also possible for response functions in the short regime. 
We use that any response function $R_{A,B}$ as defined below (\ref{equ:FDR}) 
can be written -- via, e.g., the approach of~\cite{MaySol04} -- in the form 
\begin{eqnarray}
  R_{A,B}(t,\tw) & = & \sum_k \sum_{\bsig,\bsig'} A(\bsig') \left[ 
  p_{\dt}(\bsig'|F_k \bsig) - p_{\dt}(\bsig'|\bsig) \right] \nonumber \\
  & & \times \Delta_k B(\bsig) w_k(\bsig) \left[ 1-w_k(\bsig) \right] 
  p_{\tw}(\bsig).
  \label{equ:RAB}
\end{eqnarray}
This equation applies for general systems governed by heat-bath dynamics 
with Glauber rates $w_k$ and for generic observables $A,B$. 
The $p_{\tw}(\bsig)$ 
denote state probabilities at time $\tw$ while $p_{\dt}(\bsig'|\bsig)$ are 
conditional probabilities to go from state $\bsig$ to $\bsig'$ during the 
time interval $\dt$. In (\ref{equ:RAB}), $\Delta_k B(\bsig) = B(F_k \bsig) - 
B(\bsig)$ is just a short hand expressing the change of $B$ under a 
spin-flip. 

In the concrete case of defect pair observables $A = B = n_i n_{i+1}$ we have 
$\Delta_k B(\bsig) = 0$ except for $k=i,i+1,i+2$. Denote the corresponding 
contributions to the response by $r_1,r_2,r_3$, respectively. Next work 
out $a_k = \Delta_k B(\bsig) w_k(\bsig) [1-w_k(\bsig)]$; it is convenient 
to use that Glauber rates at $T=0$ for the Ising chain may be written as 
$w_k = \frac{1}{2} (n_{k-1}+n_k)$. It turns out that $a_{i+1} = 0$ and hence 
$r_2 = 0$. Also, $r_1$ and $r_3$ are related by reflection symmetry so we only 
discuss $r_1$. From $a_i$ and (\ref{equ:RAB}), 
\begin{eqnarray*}
  r_1 & = & \frac{1}{4} \sum_{\bsig,\bsig'} n_i n_{i+1} \left[ 
  p_{\dt}(\bsig'|F_i \bsig) - p_{\dt}(\bsig'|\bsig) \right] \\
  & & \times [n_{i-1} (1-n_i) n_{i+1} - (1-n_{i-1}) n_i n_{i+1}] p_{\tw}(\bsig).
\end{eqnarray*}
In the first line of this equation $n_k$ is to be read as $n_k(\bsig')$
while in the  second one $n_k = n_k(\bsig)$. Now consider the term $n_{i-1} (1-n_i) n_{i+1}$. 
It only contributes to $r_1$ for states $\bsig$ containing defects at sites $i-1$ 
and $i+1$ but not at site $i$. For zero temperature coarsening and at large $\tw$, 
however, we have that if there are defects on sites $i-1$ and $i+1$ then 
to $O(\tw^{-3})$ there will be no further defects in any finite neighbourhood 
anyway. We also know that the density of states containing defects at sites 
$i-1, i+1$ is $\langle n_{i-1} n_{i+1} \rangle = D_2(\tw) + O\left(\tw^{-3}\right)$. 
Next, given any such state $\bsig$, $\sum_{\bsig'} n_i(\bsig') n_{i+1}(\bsig') 
p_{\dt}(\bsig'|\bsig)$ is the probability that these defects occupy sites $i,i+1$ 
a time $\dt$ later, that is $G_{\dt}(i,i+1;i-1,i+1)$ from (\ref{equ:G}). The state 
$F_i \bsig$, on the other hand, has its defects on sites $i,i+1$. [To
see this, note that $F_i$ 
is a {\em spin} flip operator $\sigma_i \to -\sigma_i$, so $n_{i-1}(F_i \bsig) = 
1-n_{i-1}(\bsig)$ and $n_i(F_i \bsig) = 1-n_i(\bsig)$ using $n_k = \frac{1}{2} 
(1-\sigma_k \sigma_{k+1})$.] Consequently $\sum_{\bsig'} n_i(\bsig') n_{i+1}(\bsig') 
p_{\dt}(\bsig'|F_i \bsig) = G_{\dt}(i,i+1;i,i+1)$. Repeating this argument for 
the term $(1-n_{i-1}) n_i n_{i+1}$ in $r_1$, where $\langle n_i n_{i+1} \rangle = 
D_1(\tw) + O\left(\tw^{-3}\right)$, and working out $r_3$ in the same fashion 
then produces 
\begin{eqnarray}
  R(t,\tw) & = & \frac{1}{2} \left[ G_{\dt}(0,1;0,1) - G_{\dt}(0,1;0,2) \right] 
  \nonumber \\
  & & \times \left[ D_1(\tw) + D_2(\tw) \right] + O\left( \tw^{-3} \right). 
  \label{equ:RGD}
\end{eqnarray}
In this result invariance of $G$ under index translation and $G_{\dt}(0,1;-1,1) = 
G_{\dt}(0,1;0,2)$ was used. From $D_1(\tw) = \widetilde{D}_1(\tw)$,  $D_2(\tw) = \widetilde{D}_2(\tw) + 
O\left( \tw^{-3} \right)$ and (\ref{equ:P}), (\ref{equ:G}) one shows that 
(\ref{equ:RGD}) precisely reproduces the short-time response $\Rst$, 
Eq.~(\ref{equ:Rst}). The subdominant $O\left( \tw^{-3} \right)$ corrections in 
(\ref{equ:RGD}) are aging contributions arising from multi-defect processes. 

The structure of (\ref{equ:RGD}) clearly reflects the mechanisms causing a 
short-time response. During the time interval $[\tw,\tw+\delta t]$
where the perturbation $\delta \mathcal{H} = -h n_i n_{i+1}$ is applied there 
is an increased likelihood for a defect pair located at sites $i,i+1$ 
to stay there. The effect on $\langle n_i(t) n_{i+1}(t) \rangle$ is
accounted for  
by $G_{\dt}(0,1;0,1) D_1(\tw)$ in (\ref{equ:RGD}). 
Conversely, the chances for the defect on site $i+1$ to move to 
$i+2$ during the interval $[\tw,\tw+\delta t]$ are decreased. 
The corresponding change in $\langle n_i(t) n_{i+1}(t) \rangle$ 
is proportional to $- G_{\dt}(0,1;0,2) D_1(\tw)$.
By the same reasoning and starting from 
configurations containing defects on sites $i,i+2$ or $i-1,i+1$ one explains 
the remaining terms in (\ref{equ:RGD}). Overall, defects are on average closer 
to each other and more likely to occupy sites $i,i+1$ at time $\tw+\delta t$ 
due to the perturbation. However, this increases the chances for subsequent 
annihilation of the defect pair so that we should expect 
$\langle n_i(t) n_{i+1}(t) \rangle$ to become lower than without the
perturbation
eventually. Indeed, from (\ref{equ:RGD}) and $D_2(\tw) \sim 2 D_1(\tw)$ 
the instantaneous response $R(\tw,\tw) \sim \frac{3}{2} D_1(\tw)$ is positive. 
But as we increase $\dt$ the response drops quickly 
and becomes zero at $\dt = \tau^* \approx 2.132$; here $\tau^*$ is 
the solution of $G_{\tau^*}(0,1;0,1) = G_{\tau^*}(0,1;0,2)$. For $\dt > 
\tau^*$ the response is negative and ultimately vanishes as $O\left(\dt^{-2}\right)$ 
in the short-time regime.

Our discussion so far explains the shape and origin of the 
short-time nonequilibrium response. But we still do not have an answer 
as to why $\Rst$ differs from its equilibrium counterpart and thus 
violates quasi-equilibrium. To the contrary, from the above reasoning 
it actually seems puzzling that we found 
$R_{\eq}(\dt,\taueq) > 0$ in equilibrium; see end of
Section~\ref{sec:eqlargetimes}. The answer to this problem 
is non-trivial: although the rate for defect pair creation 
is negligible at low temperatures, perturbations of this 
process contribute in leading order to the equilibrium response. 
For coarsening dynamics, on the other hand, such processes are absent 
[at $T=0$] or negligible [at $T>0$, see Section~\ref{sec:equilibration}]. 
Unfortunately, when using $T > 0$ rates $w_n$ in (\ref{equ:RAB}) the 
simple random walk analysis from above cannot be repeated. We therefore 
limit our discussion to the instantaneous response. 
From $p_{\dt}(\bsig'|\bsig) = \delta_{\bsig',\bsig}$ at $\dt = 0$ 
and setting $A=B$ in (\ref{equ:RAB}), 
$$
  R_{A,A}(\tw,\tw) = \sum_{k,\bsig} \left[\Delta_k A(\bsig) \right]^2 
  w_k(\bsig) \left[ 1-w_k(\bsig) \right] p_{\tw}(\bsig).
$$
Writing $w_k = \frac{1}{2}(1-\gamma)+\frac{1}{2}\gamma(n_{k-1}+n_k)$ 
and substituting $A=n_i n_{i+1}$ it is straightforward to show that the 
instantaneous defect pair response at $T > 0$ is 
\begin{equation}
  R(\tw,\tw) = \frac{1}{2\taueq}\frac{1+\gamma}{2} + \frac{1}{2} \left[ 
    D_1(\tw) + \gamma^2 \widetilde{D}_2(\tw) \right].  
  \label{equ:rescor}
\end{equation}
The first term in 
this result accounts for perturbations of the pair creation rate. 
Defect pair creation at sites $i,i+1$ corresponds to flipping spin 
$\sigma_{i+1}$ within a domain, where $\sigma_i = \sigma_{i+1} =
\sigma_{i+2}$. From the Ising Hamiltonian the associated cost in energy is 
$\Delta_{i+1} \mathcal{H}(\bsig) = 4J$. In the presence of the perturbation 
$\delta\mathcal{H} = -h n_i n_{i+1}$, however, this 
is lowered to $4J-h$. Therefore, at $T > 0$, the perturbation increases the 
rate $w_{i+1}$ of such spin flips and thus the density of defect pairs;
recall that in the limit of low temperatures we first take $h \to 0$ 
and then $T \to 0$ so that the calculated response is always linear.
From (\ref{equ:rescor}) and at low temperatures, where $\gamma \sim 1$,  
this produces a contribution of $1/(2\taueq)$ in the instantaneous response. 
Now compare this to the other terms in (\ref{equ:rescor}), using that 
$\widetilde{D}_2 \sim D_2$ at low $T$ whether in or out of equilibrium. 
For low temperature equilibrium we have $D_{2,\eq} \sim D_{1,\eq}$. 
Because $D_{1,\eq} \sim 1/(2\taueq)$ we may write $R_{\eq} \sim 2 D_{1,\eq}$, 
with all terms in (\ref{equ:rescor}) being of the same order. 
Out of equilibrium, on the other hand, $D_2 \sim 2 D_1$. 
The first term in (\ref{equ:rescor}) is absent at $T=0$ or negligible compared 
to the others at $T>0$ for sufficiently small $\tw$ [see
Section~\ref{sec:equilibration}]. Overall, we thus have the
instantaneous response $R \sim \frac{3}{2} D_1$ for coarsening but 
$R_{\eq} \sim 2 D_{1,\eq}$ in equilibrium. It is this difference in
the prefactors that leads to $X(\tw,\tw) = \frac{3}{4}$.

\subsection{FD Limit Plot}
\label{sec:FDplot}

\begin{figure}
  \epsfig{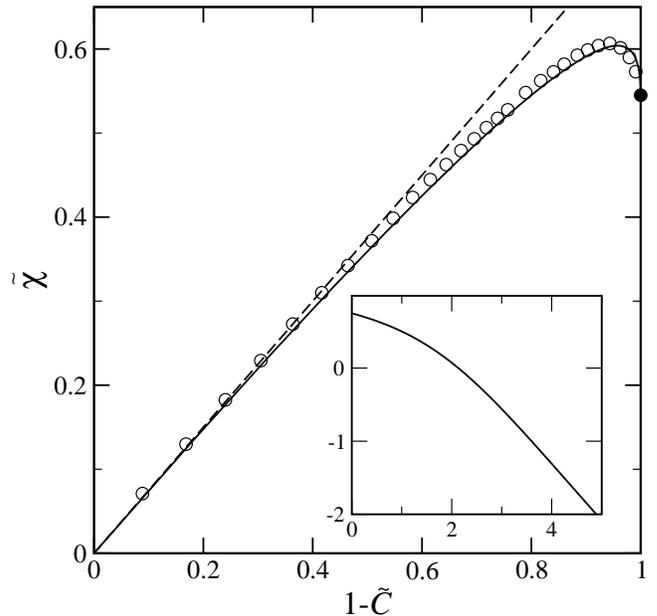} 
  \caption{\label{fig:FDplot} FD-plots for the defect pair
  observable $A_1 = n_i n_{i+1}$. The full curve 
  shows the limit plot defined by
  (\ref{equ:cortilde}), (\ref{equ:chitilde}). The dashed line 
  has slope $X=\frac{3}{4}$ and
  is tangent to the limit plot at the origin. 
  The black dot on the right marks the end-point $(1,Y_1)$ of 
  the limit plot. Simulation data are represented by circles. 
  They are obtained from simulations at $T=0$ in 
  a system of $10^6$ spins. The probing field for the susceptibility is 
  $\bar{h} = 0.5$ and the data are averaged over 500 runs. In the plot,
  $\tw = 10$ is fixed and $t = 10 \ldots 15$ is the running parameter. 
  Inset: A plot of the FDR in the short-time regime $X(\dt)$, 
  Eq.~(\ref{equ:Xqe}), versus $\dt$.}
\end{figure}

From (\ref{equ:coragdtO1}), (\ref{equ:resagdtO1}) we 
have that the two-time functions $C(t,\tw)$, $R(t,\tw)$ 
drop to an arbitrarily small fraction of their equal time
values within the short-time regime, i.e.\ before they 
display aging. Therefore the exact
FD-limit plot follows from the short-time expansions. 
Since the amplitudes of equal time
quantities are time dependent we normalize 
$\widetilde{\chi}(t,\tw) = \chi(t,\tw)/C(t,t)$
and $\widetilde{C}(t,\tw) = C(t,\tw)/C(t,t)$ 
and plot $\widetilde{\chi}$ against $1-\widetilde{C}$,  
see~\cite{SolFieMay02,MayBerGarSol04}. From 
(\ref{equ:coragdtO1}), (\ref{equ:resagdtO1}) one 
obtains 
\begin{eqnarray}
  \widetilde{C} & = & p_1(\dt), 
    \label{equ:cortilde} \\
  \widetilde{\chi} & = & \frac{3}{4} \left[ \left( 1 - p_1(\dt) \right) 
    - 2 \int_0^{\dt} \dd\tau \me^{-2\tau} \frac{I_1^2(\tau)}{\tau} \right]. 
    \label{equ:chitilde}
\end{eqnarray}
These equations apply in the limit $\tw \to \infty$ for arbitrary
fixed  $\dt \geq 0$.
The resulting FD-plot is shown in Figure~\ref{fig:FDplot}. Note that 
when constructing FD-plots one generally has to keep $t$ fixed and use $\tw$ 
as the curve parameter~\cite{SolFieMay02}. This convention ensures that the 
slope of the FD-plot is $X(t,\tw)$. In the short-time regime we are exploring, 
however, the normalized functions only depend on $t-\tw$ and either
$t$ or $\tw$ may be used as the plot parameter. This is exact for 
$\tw \to \infty$ and correct to leading order in $\tw$ at finite $\tw$.

In Fig.~\ref{fig:FDplot} the slope of the plot at the origin, 
where $\tw=t$, is given by $X(0)=\frac{3}{4}$.  As $\dt$ increases
and reaches $\tau^*$ the response goes to zero.  Consequently the
susceptibility reaches a maximum at $\dt=\tau^*$ and the tangent to
the FD-plot in Figure~\ref{fig:FDplot} becomes horizontal, with
$X(\tau^*)=0$.  As 
we increase $\dt$ further the FDR (\ref{equ:Xqe}) turns negative and
diverges linearly with $\dt$, $X(\dt) \to -\infty$. Hence the FD-plot
becomes vertical 
as it approaches its end point $(1,Y_1)$. Taking $\dt \to \infty$ in 
(\ref{equ:chitilde}), where the integral is solvable~\cite{Mathbook}, 
gives $Y_1 = \frac{3}{2} - \frac{3}{\pi} \approx 0.545$. 
Fluctuation dissipation relations for the
aging case, where $\dt$ and $\tw$ are comparable, are compressed into
this point. So the plot in Figure~\ref{fig:FDplot} only reflects the 
fluctuation dissipation behaviour in the short-time regime. In order 
to demonstrate that the predicted violation of quasi-equilibrium can 
easily be observed in simulations we have included such data in 
Fig.~\ref{fig:FDplot}.

\subsection{Beyond Short Time Differences}
\label{sec:longtime}

Our discussion of the non-equilibrium coarsening dynamics so far was
focused on the short-time regime where $\dt \geq 0$ finite and $\tw \to 
\infty$; only the short-time terms $\Cst, \Rst$ in our expressions 
(\ref{equ:Csa}),(\ref{equ:Rsa}) for the connected two-time defect pair 
correlation and response functions contributed to leading order in 
$\tw$. Let us now briefly summarize some interesting features of 
$C(t,\tw)$ and $R(t,\tw)$ beyond the short-time regime.  Here 
$\dt, \tw \to \infty$ simultaneously, and therefore the aging 
contributions $\Cag, \Rag$, Eqs.~(\ref{equ:Cag}), ({\ref{equ:Rag}}), 
must be taken into account. 

\begin{figure}
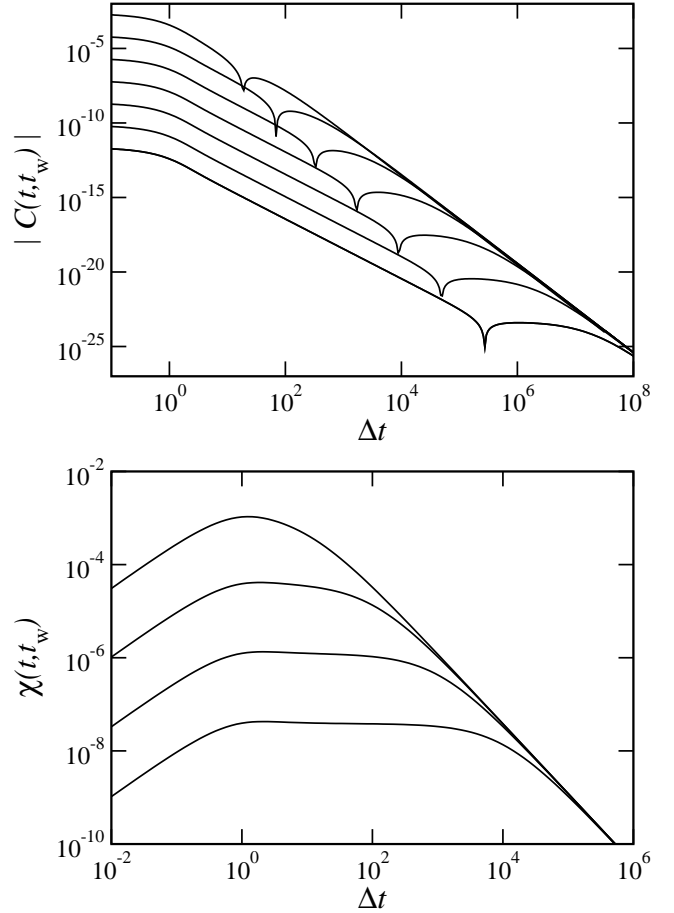

  \epsfig{file=clongtime.eps, width=8.5cm, clip} \\[1ex]
  \epsfig{file=slongtime.eps, width=8.5cm, clip} 
  \caption{\label{fig:longtime} Top: Plots of the exact connected two-time 
  defect pair correlation function $C(t,\tw)$ for zero temperature coarsening 
  dynamics, obtained from (\ref{equ:Csa}), 
  (\ref{equ:Cst}), (\ref{equ:Cag}). The curves correspond to $\tw = 10^1, 
  10^2,\ldots ,10^7$ from top to bottom, respectively. The cusp in each 
  curve separates the region $C(t,\tw) > 0$ [at small $\dt$] from $C(t,\tw) 
  < 0$ [at large $\dt$]. Bottom: Plots of the exact two-time defect pair 
  susceptibility $\chi(t,\tw)$ obtained by numerical integration from 
  (\ref{equ:Rsa}), (\ref{equ:Rst}), (\ref{equ:Rag}). The curves correspond 
  to $\tw = 10^1, 10^2, 10^3, 10^4$ from top to bottom, respectively.}
\end{figure}

For correlations we expect to see an effect from the competition
between the pair return probability $p_1(\dt)$ and the chance of
finding an independent pair at sites $i,i+1$ at the later time $t$, 
by analogy with the situation in equilibrium; see
Section~\ref{sec:eqlargetimes}. In non-equilibrium and for small $\dt$
the disconnected and connected correlations coincide to leading 
order in $\tw$. So from (\ref{equ:coragdtO1}) the disconnected correlation 
is $C^{\mathrm{DC}}(t,\tw) \approx
D_1(\tw) p_1(t-\tw)$. Assuming that this equation applies up to
sufficiently large $\dt$ -- though still much smaller than $\tw$ -- we
may now estimate the time-scale at which competition sets in. This is
done by comparing $C^{\mathrm{DC}}(t,\tw)$ to $D_1(\tw) D_1(t)$, which
is the product of the independent probabilities of having a defect
pair at sites $i,i+1$ at time $\tw$ and at time $t$. Because we are
assuming $\dt \ll \tw$, $D_1(\tw) D_1(t)\approx D_1^2(\tw)$. The
scalings $p_1(\dt) = O\big(\dt^{-2}\big)$ and $D_1(\tw) =
O\left({\tw}^{-3/2}\right)$ then show that $C^{\mathrm{DC}}(t,\tw)$ becomes
comparable to $D_1(\tw) D_1(t)$ on the non-trivial time scale $\dt =
O\left({\tw}^{3/4}\right)$. In fact, from the plots in Fig.~\ref{fig:longtime} 
the connected correlation
becomes {\em negative} on that time scale. This means that for $\dt \gg
{\tw}^{3/4}$, the chances of finding a defect pair at sites $i,i+1$ at
time $t$ and at time $\tw$ are {\em lower} than those of independently
finding pairs at both times: the presence of a defect pair at time $t$
is negatively correlated with that at $\tw$.

We may picture this effect as follows. If we know that there is a
defect pair at sites $i,i+1$ at time $\tw$, the neighboring defects
are likely to be at a distance of the order of the typical domain
size. Then, as time evolves, the original pair becomes more and more
likely to have disappeared via annihilation while neighboring defects
have not yet had enough time to reach sites $i,i+1$. For the
equilibrium domain size distribution these effects reach a balance on
the time-scale $\dt = O(\sqrt{\taueq})$.  In the coarsening case,
however, the relative concentration of small domains -- as compared 
to typical domains -- is much lower than in
equilibrium, so that annihilation of the original pair is
comparatively the stronger effect. Thus, on the time-scale $\dt =
O\left({\tw}^{3/4}\right)$, we have a ``hole" in the spatial distribution
of defect pairs around sites $i,i+1$. This hole persists up to the
time-scale $\dt = O(\tw)$, where neighboring defects have had time to
diffuse in eventually. The connected correlation function, see 
Fig.~\ref{fig:longtime}, therefore has three dynamical regimes at 
large $\tw$: Up to times $\dt \ll {\tw}^{3/4}$ the expansion for the 
short-time regime (\ref{equ:coragdtO1}) applies and $C(t,\tw) \sim 
\Cst(t,\tw)$. In the time window ${\tw}^{3/4} \ll \dt \ll \tw$ 
the connected correlation is negative and $t$-independent, with 
contributions from $\Cst$ negligible so that $C(t,\tw) \sim - D_1^2(\tw) 
\sim -(1/64) \pi^{-1} \tw^{-3}$. 
Finally, at large $\dt \gg \tw$ the connected correlation remains negative 
but vanishes as $C(t,\tw) \sim -(1/8) \pi^{-1} \dt^{-3}$, as follows from 
expansions of (\ref{equ:Cst}), (\ref{equ:Cag}).

Comparison of the correlation functions in Fig.~\ref{fig:dat_sim} and 
Fig.~\ref{fig:longtime} shows that the simulation data at $\tw = 1$ has 
only given us a glimpse on the full aging behaviour of $C(t,\tw)$. 
From the scales of the plot in Fig.~\ref{fig:longtime}, on the other 
hand, it is clear that such data are out of reach for simulations. 
Consider, e.g., the curve for $\tw = 10^3$: the connected correlation 
$C(t,\tw)$ drops from its equal-time value $C(\tw,\tw) \sim D_1(\tw)$ 
of about $10^{-6}$ to around $10^{-12}$, that is by six orders of magnitude, 
before it deviates from its short-time behaviour $\Cst$ and displays aging. 
This illustrates the general problem associated with exploring the 
aging behaviour of correlation functions with a scaling of the form 
(\ref{equ:CAstag}). We have discussed this issue in the context of 
single defect observables in the 1$d$ and 2$d$ Ising models 
in~\cite{MayBerGarSol03}. There, long-time FD-plots are trivial with 
$\widetilde{\chi} = 1 - \widetilde{C}$. However, this only reflects that 
quasi-equilibrium is satisfied and does not reveal any information 
about the aging regime. A similar situation is encountered in the 
1$d$ FA model which, despite a trivial FD-plot~\cite{BuhGar02b,Buhot03}, 
has $X \neq 1$ in the aging regime~\cite{MayBerGarSol05}. As regards the 
issue of measuring the asymptotic FDR
$X^\infty = \lim_{\tw\to\infty}\lim_{t\to\infty} X(t,\tw)$, a solution
for this problem was suggested in~\cite{MayBerGarSol03}.
It consists in using different observables which share the same 
$X^\infty$ but are more easily accessed in simulations. 

The aging behaviour of the two-time defect pair response function 
$R(t,\tw)$ is rather simple. Analysis of (\ref{equ:Rst}), (\ref{equ:Rag}) 
shows that the short-time expansion $R(t,\tw) \sim \Rst(t,\tw)$, 
Eq.~(\ref{equ:resagdtO1}), applies until $\dt$ becomes compareable to 
$\tw$. More precisely, for $1 \ll \dt \ll \tw$, we have $R(t,\tw) \sim 
-(3/32)\pi^{-3/2} \dt^{-2} {\tw}^{-3/2}$. Note that as discussed in 
Sec.~\ref{sec:response} the response is negative. 
In the opposite limit $1 \ll \tw \ll \dt$ this crosses over to 
$R(t,\tw) \sim -(9/32)\pi^{-3/2} \dt^{-3} {\tw}^{-1/2}$, accelerating 
the decrease of $R(t,\tw)$ by a factor of $\tw/\dt$. Intuitively 
speaking, the two defects that were located near sites $i,i+1$ at time 
$\tw$ and caused the response are likely to have annihilated with 
other defects in the system when $\dt \gg \tw$. This decreases the chances 
for such a defect pair to return to sites $i,i+1$, and therefore the 
response.

The scaling of $R(t,\tw)$ has an interesting consequence 
for the susceptibility $\chi(t,\tw) = \int_{\tw}^t d\tau R(t,\tau)$, 
also referred to as zero-field-cooled (ZFC) susceptibility. At large 
$t$ this integral is dominated by the short-time response $\Rst$, 
i.e.\ $\chi(t,\tw) \sim \Sst(t,\tw)$.  
In the integral, as $\tau$ departs from $t$ the modulus of the integrand 
$R(t,\tau)$ drops like $(t-\tau)^{-2}$ for $t-\tau \gg 1$ and therefore 
the integral converges quickly. Aging effects in the scaling of $R(t,\tau)$ 
when $t-\tau \gg \tau$, i.e.\ $\tau \ll t/2$, only give small
corrections to 
the value of the integral. Therefore aging contributions in the ZFC 
susceptibility $\chi(t,\tw)$ are {\em subdominant}. At large $t$ and 
for any $\tw \leq t$ this implies $\chi(t,\tw) \sim D_1(t) 
\widetilde{\chi}(t-\tw)$ with $\widetilde{\chi}$ as given by 
(\ref{equ:chitilde}). Consequently we have $\chi(t,\tw) \sim Y_1 D_1(t)$ 
in the aging regime, see Fig.~\ref{fig:longtime}, and contributions from 
aging effects in the response have vanished in $\chi(t,\tw)$.
We can see the extent of this 
effect by looking back to Fig.~\ref{fig:dat_sim}: already at the very
moderate value of $\tw = 1$, aging contributions 
in $\chi(t,\tw)$ are marginal. While all of this is immaterial for exact 
calculations, where we start from the response in the first place, it is 
crucial for interpreting simulation results. In problems where the 
response function has a scaling analogous to (\ref{equ:CAstag}), 
see~\cite{MayBerGarSol03,BuhGar02b,Buhot03}, 
measurement of the ZFC susceptibility gives $\chi(t,\tw) \sim \Sst(t,\tw)$ 
and no information about the aging behaviour of $R(t,\tw)$ can be 
extracted. In a measurement of the so-called thermoremanent (TRM) 
susceptibility $\rho(t,\tw) = \int_0^{\tw} d\tau R(t,\tau)$, 
on the other hand, this bias is not present since $t-\tau \geq t-\tw$. 
So if $\dt > \tw$, for example, the integral only contains contributions 
from $R(t,\tau)$ with $t-\tau > \tau$ and aging in $R$ is revealed. 
The situation is precisely reversed as compared to the case of 
spin observables in critical coarsening~\cite{CorLipZan03}, where the aging 
behaviour of $R$ can be extracted from $\chi$ but not from $\rho$. 

The nonequilibrium FDR $X(t,\tw)$ as obtained from (\ref{equ:Csa})-(\ref{equ:Rag}) 
has rather strange features when $\dt$ and $\tw$ are simultaneously large. 
But since the observable $A_1$ does not produce quasi-equilibrium FDT in the 
short-time regime we do not expect $X(t,\tw)$ to have a sensible meaning 
in the context of effective temperatures. We comment only that the 
short-time expansion (\ref{equ:Xqe}) applies as long as $\dt \ll
{\tw}^{3/4}$, while the asymptotic FDR diverges, $X^\infty = \infty$.

\subsection{Equilibration}
\label{sec:equilibration}

\begin{figure}
  \epsfig{file=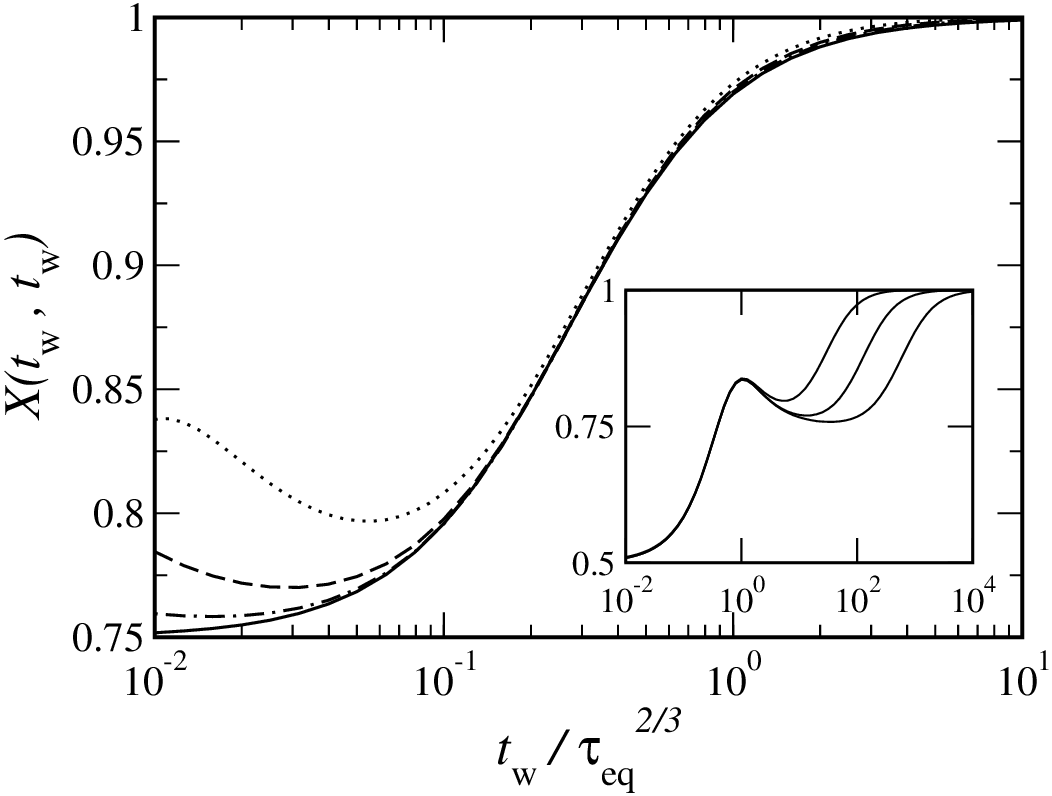, width=8.5cm, clip} 
  \caption{\label{fig:Xtwtw} Plots of the equal time FDR $X(\tw,\tw)$ 
  versus rescaled time $\tw/{\taueq}^{2/3}$ for three different temperatures. 
  The curves correspond to $\taueq=10^3$ (dotted), $\taueq=10^4$ 
  (dashed) and $\taueq=10^5$ (dashed-dotted). The full curve applies in 
  the scaling limit $\tw,\taueq\to\infty$ with $\tw/{\taueq}^{2/3}$ fixed. 
  Inset: Plots of 
  $X(\tw,\tw)$ versus $\tw$ for $\taueq=10^3,10^4,10^5$. The hump at 
  $\tw=1$ is caused by transients following the quench.}
\end{figure}

Let us finally consider in more detail the crossover to equilibrium
for a finite temperature quench $T>0$. We focus again on the
short-time regime and, for simplicity,
discuss only the equal time FDR $X(\tw,\tw)$. An expression for
$(\partial/\partial \tw) C(t,\tw)|_{t=\tw}$ is obtained from
(\ref{equ:cor2t}). The instantaneous response $R(\tw,\tw)$ 
follows most conveniently from (\ref{equ:rescor}) by working 
out $\widetilde{D}_1(\tw)$ and $\widetilde{D}_2(\tw)$ for a quench to $T > 0$~\cite{MaySol04}. 
The $\tw$-dependence of
the resulting $X(\tw,\tw)$ for three different temperatures is shown
in Figure~\ref{fig:Xtwtw}. As expected the curves cross over from
$X(\tw,\tw)=\frac{3}{4}$ at sufficiently small $\tw$ [but still 
$\tw \gg 1$] to the
equilibrium value $X(\tw,\tw)=1$ at large $\tw$. The time scale for
equilibration of $X(\tw,\tw)$, however, is set by ${\taueq}^{2/3}$.  In
order to understand this result we consider the densities of small
domains: in equilibrium we have the scaling $D_{1,\eq}(\taueq) \sim
1/(2\taueq)$ whereas for zero temperature coarsening $D_1(\tw) \sim
1/\left(8 \, \sqrt{\pi} \, {\tw}^{3/2}\right)$. This shows that the
dynamical density $D_1(\tw)$ is comparable to the equilibrium density
$D_{1,\eq}(\taueq)$ for $\tw = O\left({\taueq}^{2/3}\right)$.  By working
out $D_1 = \langle n_i n_{i+1} \rangle$ for a quench to $T > 0$,
one easily
verifies that $D_1$ indeed becomes stationary at its equilibrium value
for $\tw \gg {\taueq}^{2/3}$. In this regime one also finds $D_2\sim
D_1$ as expected for low-$T$ equilibrium, rather than $D_2\sim 2 D_1$
in the coarsening regime. From the representation (\ref{equ:rescor})
of $R(\tw,\tw)$ it then follows that the process of defect pair
creation starts to contribute to the instantaneous response for $\tw
\gg {\taueq}^{2/3}$, and that the ratio of the other terms assumes its
equilibrium value. It is then not surprising that also the 
$\dt$-dependence of $R(t,\tw)$ becomes identical to the equilibrium 
response throughout the short-time regime, i.e., quasi-equilibrium 
behavior is recovered for $\tw \gg {\taueq}^{2/3}$. 

\section{Non-adjacent defects}
\label{sec:multi}

In the previous two sections we presented a comprehensive discussion 
of the functions $C(t,\tw) \equiv C_1(t,\tw)$ and $R(t,\tw) \equiv R_1(t,\tw)$ 
for the observable $A_1 = n_i n_{i+1}$. Now we investigate to which 
extent our findings generalize to observables $A_d = n_i n_{i+d}$
which detect 
defects a distance $d$ apart. As explained in the Appendix, an exact derivation of the associated functions 
$C_d(t,\tw)$ and $R_d(t,\tw)$ as defined in (\ref{equ:cor2tdef}) and 
(\ref{equ:res2tdef}), respectively, would be extremely cumbersome.
Instead we exploit the fact that, in the short-time regime we are
interested in, these functions can to leading order in $\tw$ be
obtained from random walk arguments.

Consider the states $\bsig$ obtained from low temperature coarsening dynamics 
at large $\tw$. In complete analogy to the case $d=1$ we have that amongst the 
states $\bsig$ containing defects at sites $i$ and $i+d$ only a fraction 
$O\left( \tw^{-3} \right)$ have further defects in any finite neighbourhood. 
Therefore the discussion of (\ref{equ:coreqdtO1}) or (\ref{equ:coragdtO1}) 
directly generalizes to any finite $d > 1$. In terms of the probability for a 
defect pair initially located at sites $i,i+d$ to occupy the same sites a time 
$\dt$ later, 
\begin{equation}
  p_d(\dt) = G_{\dt}(i,i+d;i,i+d), 
  \label{equ:pd}
\end{equation}
the correlaton $C_d(t,\tw)$ in the short-time regime is 
\begin{equation}
  C_d(t,\tw) = p_d(\dt) D_d(\tw) + O\left( \tw^{-3} \right). 
  \label{equ:Cd}
\end{equation}
Although this argument applies directly to the disconnected correlations it 
is also true for $C_d(t,\tw)$ since both 
agree to $O\left(D_d^2\right) = O\left(\tw^{-3}\right)$ in the short 
time regime. For response functions $R_d(t,\tw)$ we use (\ref{equ:RAB}) 
and the same reasoning as in Section~\ref{sec:response} to 
to obtain for $d>1$
\begin{eqnarray}
  R_d & = & + \frac{1}{2} \left[ G_{\dt}(0,d;0,d) - G_{\dt}(0,d;0,d-1) \right] D_{d-1}  
  \nonumber \\ 
  & & + \frac{1}{2} \left[ G_{\dt}(0,d;0,d) - G_{\dt}(0,d;0,d-1) \right] D_d   
  \nonumber \\ 
  & & + \frac{1}{2} \left[ G_{\dt}(0,d;0,d) - G_{\dt}(0,d;0,d+1) \right] D_d   
  \nonumber \\ 
  & & + \frac{1}{2} \left[ G_{\dt}(0,d;0,d) - G_{\dt}(0,d;0,d+1) \right] D_{d+1} 
  \nonumber \\
  & & + O\left( \tw^{-3} \right).
  \label{equ:Rd} 
\end{eqnarray}
[In order to save space we have omitted here the time arguments $R_d =
R_d(t,\tw)$ and $D_d = D_d(\tw)$ etc.] An expression for the 
$G$'s in (\ref{equ:Cd},\ref{equ:Rd}) is stated in (\ref{equ:G}),
giving in particular 
$p_d(\dt) = \me^{-2\dt} \left[ I_0^2 - I_d^2 \right](\dt)$, while we
can estimate $D_d(\tw) = \widetilde{D}_d(\tw) + O\left(\tw^{-3}\right)$ using
(\ref{equ:P}).

Before we proceed with a discussion of (\ref{equ:Cd}), (\ref{equ:Rd}) for 
nonequilibrium coarsening dynamics, let us briefly consider an equilibrium 
situation. The above assumption regarding the nature of the states $\bsig$ that contribute to $C_d$ and 
$R_d$ then still applies if the temperature is low. Therefore, to leading 
order in $\taueq$, the equivalent of (\ref{equ:coreqdtO1}) for $d > 1$ is 
$C_{d,\eq}(\dt,\taueq) \sim p_d(\dt) D_{d,\eq}(\taueq)$ from (\ref{equ:Cd}). 
In (\ref{equ:Rd}), on the other hand, 
we use $D_{d \pm 1,\eq}(\taueq) \sim D_{d,\eq}(\taueq)$ 
as the density of small domains is flat in low $T$ equilibrium. Combining 
terms then shows that $R_{d,\eq}(\dt,\taueq) \sim [(-\partial/\partial \dt) p_d(\dt)] 
D_{d,\eq}(\taueq)$. Thus equilibrium FDT is recovered from (\ref{equ:Cd}), 
(\ref{equ:Rd}) at low temperatures. This is non-trivial because we use 
zero temperature Glauber rates $w_n$ in the derivation of (\ref{equ:Rd}), see 
Section~\ref{sec:response}. In contrast to the $d=1$ case of adjacent defects, 
pair creation processes do {\em not} contribute in leading order to
the responses  
$R_d$ with $d > 1$. This makes sense: the perturbation $\delta\mathcal{H} = 
-h n_i n_{i+d}$ acts on sites a distance $d$ apart but pair creation is only 
possible on adjacent sites. So the pair creation 
rate for sites $i+d,i+d+1$ [say] is affected only if we already have a
defect present at site $i$. The latter condition makes such
contributions in $R_d$ subdominant for $d \geq 2$.

\begin{figure}
  \epsfig{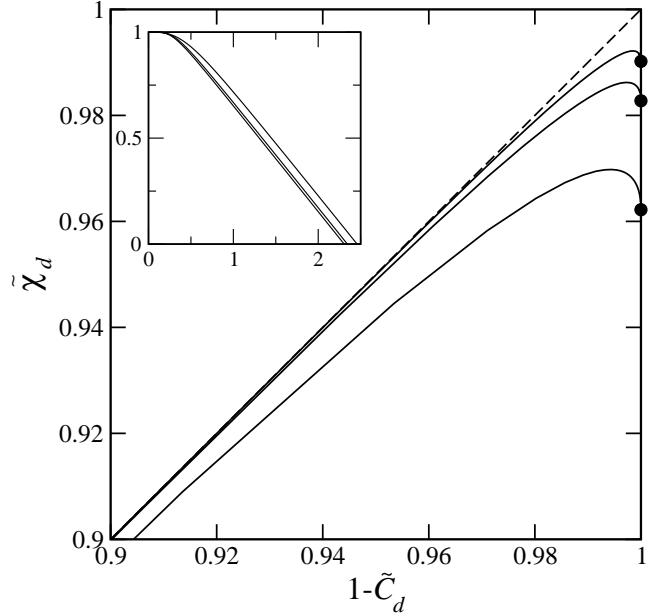} 
  \caption{\label{fig:FDPdgt2} FD limit plots for defect pair
  observables $A_d = n_i n_{i+d}$ with $d=2,3,4$ from bottom to 
  top, respectively, obtained from 
  (\ref{equ:cordtilde}), (\ref{equ:chidtilde}). The dashed line   
  has slope $X=1$ and represents equilibrium FDT. 
  Black dots mark the end-points $(1,Y_d)$ of 
  the limit plots. 
  Inset: Plots of the FDRs in the short-time regime $X_d(\dt)$, Eq.~(\ref{equ:Xd}), 
  with $d=2,3,4$ from top to bottom, respectively, versus $\dt/d^2$. 
  The curves for $d=3,4$ are almost indistinguishable, showing that $X_d$ quickly 
  approaches a scaling form.}
\end{figure}

Having clarified this qualitative difference between the responses $R_1$ and 
$R_d$ with $d \geq 2$ we return to nonequilibrium coarsening dynamics. Here 
the density of small domains $D_d(\tw)$ is to leading order proportional to 
the domain size $d$; more precisely we estimate 
$D_{d \pm 1}(\tw)$ using (\ref{equ:P}) and $\me^{-x} I_{d \pm 1}(x) 
= \me^{-x} I_d(x) + O\left( x^{-3/2} \right)$. This allows us to rearrange 
(\ref{equ:Rd}) into 
\begin{eqnarray}
  R_d(t,\tw) & = & \left[ \frac{\partial}{\partial \tw} 
  p_d(t-\tw) - \me^{-2(t-\tw)} \frac{I_d^2(t-\tw)}{t-\tw} \right] \nonumber \\ 
  & & \times  D_d(\tw) + O\left( \tw^{-5/2} \right). 
  \label{equ:Rdag}
\end{eqnarray}
Clearly the nonuniform density of small domains produces a nonequilibrium 
term in $R_d$. It therefore differs from its short-time equilibrium form. 
From (\ref{equ:Cd}), (\ref{equ:Rdag}) we obtain for the associated FDR in 
the short-time regime, again abbreviating $X_d(\dt) = \lim_{\tw \to \infty} 
X_d(t,\tw)$, 

\begin{equation}
  X_d(\dt) = 1 - \frac{1}{(-\partial/\partial \dt) p_d(\dt)}
  \me^{-2\dt} \frac{I_d^2(\dt)}{\dt}.
  \label{equ:Xd}
\end{equation}
The equations (\ref{equ:Cd}), (\ref{equ:Rdag}), (\ref{equ:Xd}) have a 
surprising structural similarity with their $d = 1$ counter-parts 
(\ref{equ:coragdtO1}), (\ref{equ:resagdtO1}), (\ref{equ:Xqe}). 
But, in contrast to the case of adjacent defect pairs, $X_d(0) = 1$ for 
$d > 2$ rather than $X_1(0) = \frac{3}{4}$, Eq.~\ref{equ:X34}. 
From the discussion above we see that effects from perturbing the 
defect pair creation rate, which are subdominant when $d \geq 2$ but 
contribute to leading order for $d = 1$, are at the origin of this 
difference. However, it should be stressed that for $\dt > 0$ all FDRs
$X_d(\dt)$ deviate from 
unity on an $O(1)$ time-scale. Therefore there is no quasi-equilibrium 
regime for any defect pair observable $A_d$. Instead 
we have from (\ref{equ:Xd}) that $X_d(\dt) = 1 - O\left(\dt^{2d-1}\right)$ 
for $\dt \ll d^2$ while $X_d(\dt) = -\dt/(2 d^2) + O(1)$ for $\dt \gg d^2$, 
see Fig.~\ref{fig:FDPdgt2}. 

We visualize the violation of quasi-equilibrium behaviour in terms of 
FD-plots. The scalings (\ref{equ:Cd}) and (\ref{equ:Rd}) show that with 
increasing $\tw$ plots become progressively dominated by the short 
time behaviour of $C_d$ and $R_d$. A parameterization of the limit plots 
is obtained by taking $\tw \to \infty$ at fixed $\dt$. Normalizing 
correlations and susceptibilities as in Section~\ref{sec:FDplot} gives 
\begin{eqnarray}
  \widetilde{C}_d & = & p_d(\dt), 
    \label{equ:cordtilde} \\
  \widetilde{\chi}_d & = & \left( 1 - p_d(\dt) \right) 
    - \int_0^{\dt} \dd\tau \me^{-2\tau} \frac{I_d^2(\tau)}{\tau}. 
    \label{equ:chidtilde}
\end{eqnarray}
The limit-plots for $d=2,3,4$ are presented in Fig.~\ref{fig:FDPdgt2}. 
Each plot has slope $X_d(0) = 1$ at the origin [not shown] and follows the 
equilibrium line rather closely until $\dt \approx 0.25 d^2$. Somewhere in 
the range $2 d^2 < \dt < 2.5 d^2$ the plots reach a maximum where
$X_d=0$, and acquire a vertical tangent as they approach
their end points $(1,Y_d)$, with $X_d(\dt) \to -\infty$ 
diverging linearly with $\dt$. From (\ref{equ:chidtilde}) the $Y_d$ 
are~\cite{Mathbook}  
\begin{equation}
  Y_d = 1-\frac{2}{\pi d} (-1)^d \left[ \frac{\pi}{4} + \sum_{k=1}^d 
  \frac{(-1)^k}{2k-1} \right], 
  \label{equ:Yd}
\end{equation}
giving $Y_2 \approx 0.962, Y_3 \approx 0.983, Y_4 \approx 0.990$ etc.,
and $Y_d=1-O\left(d^{-2}\right)$ for large $d$. We remark that because the 
limit plots for $d \geq 2$ lie very close to the equilibrium line -- 
Fig.~\ref{fig:FDPdgt2} shows only the top-right corner of the plot -- 
very accurate data would be needed to reproduce them in simulations. 
Furthermore $\widetilde{\chi}_d(t,\tw)$ only converges slowly to its 
$\tw \to \infty$ limit (\ref{equ:chidtilde}). This is easily verified 
by numerical evaluation of $\widetilde{\chi}_d(t,\tw) = (1/C(t,t)) 
\int_{\tw}^t \dd\tau R_d(t,\tau)$ based on (\ref{equ:Cd}), (\ref{equ:Rd}). 
These two facts in combination make it virtually impossible to see the 
{\em limit} plots, Fig.~\ref{fig:FDPdgt2}, in simulations. Equations 
(\ref{equ:Cd}) and (\ref{equ:Rd}), however, perfectly reproduce simulation 
data for $C_d(t,\tw)$ and $\chi_d(t,\tw) = \int_{\tw}^t \dd\tau R_d(t,\tau)$ 
already at small times, e.g., $\tw=10$ and $t=10\ldots 15$ as used in 
Section~\ref{sec:FDplot}. 

Our simple random walk analysis does not allow us to make predictions on 
the aging behaviour of $C_d$ and $R_d$. If both $\dt$ and $\tw$ are 
large, complicated multi-defect processes must be taken into account; 
only a calculation as sketched in the Appendix for the case $d=1$ 
would allow one to study this regime. As regards the susceptibility 
$\chi_d$, however, we can predict that $\chi_d(t,\tw) \sim Y_d D_d(t)$ in the 
aging regime even though we do not know the precise aging behaviour of 
$R_d$. This is simply because $\chi_d$ is dominated by the short-time 
response as discussed in Section~\ref{sec:longtime}. 

Finally, for quenches to small but nonzero temperature quasi-equilibrium 
behaviour will be recovered when $\tw \gg {\taueq}^{2/3}$, just as for adjacent 
defects. This follows from the dependence of the short-time response 
(\ref{equ:Rd}) on the density of small domains $D_d$ and $D_{d \pm 1}$ 
and the fact that these densities level off towards their equilibrium
values on that time-scale. 

\section{Conclusions}
\label{sec:conclusions}

In this paper we have analyzed the FDT behavior for defect-pair 
observables $A_d=n_i n_{i+d}$ in the Glauber-Ising chain.
Contrary to the commonly held notion that short-time relaxation generally 
proceeds as if in equilibrium, none of these observables produce $X=1$ 
in the short-time regime $\dt\ll\tw$; this applies as long as $\tw$ is
below the crossover timescale ${\taueq}^{2/3}$. We showed explicitly that this 
unusual behavior arises from the response functions, while the short-time
decay of correlations does indeed have an equilibrium form
apart from the expected overall amplitude factor. The deviations of
the responses from quasi-equilibrium behavior could be traced to two
factors. First, in the out-of-equilibrium response for adjacent defects,
events where pairs of domain walls are created are negligible, while in 
equilibrium they contribute at leading order. Second, all responses 
are sensitive to details of the domain-size distribution in the system,
via their dependence on the density of small domains, and these details 
differ between the equilibrium and out-of-equilibrium situations.

The inherent-structure picture mentioned in the introduction suggests 
a generic interpretation of our results: 
starting from an out-of-equilibrium configuration with a given number 
of defects or domain walls, we can loosely say that we remain within the same 
``basin'' as long as no domain walls annihilate; the energy then remains
constant. A similar interpretation has been advocated, in the context
of the Fredrickson-Andersen model, in~\cite{BerGar03b}. Transitions
to a basin with lower energy then correspond to the annihilation of
two domain walls; at long coarsening times $\tw$, such transitions
between basins are separated by long stretches of ``intra-basin''
motion.  Within this picture, our defect-pair observables $A_d$ measure
precisely when a transition to a new basin is about to happen, i.e., 
they focus on the out-of-equilibrium, inter-basin dynamics. From this 
point of view it is not surprising that the $A_d$ do not exhibit
quasi-equilibrium behavior even at short times. Spin and
single-defect observables, on the other hand, are not unusually 
sensitive to transitions between basins, so that their short-time 
relaxation is governed by the quasi-equilibrium, intra-basin 
motion~\cite{MayBerGarSol03,BuhGar02b,Buhot03}. This highlights the 
crucial dependence of nonequilibrium fluctuation-dissipation relations 
on the probing observables.

The above interpretation suggests that lack of quasi-equilibrium
behavior in the short-time regime could occur quite generically in glassy
systems.
Certainly in glass models with kinetically constrained
dynamics~\cite{RitSol03}, one 
would expect observables that detect the proximity of two or more facilitating
defects to display similar behaviour to the one studied in this paper.
More generally, the same should apply to
observables which are sensitive to transitions between 
basins or metastable states. In structural glasses, conventional observables 
such as density fluctuations are clearly not 
of this type. However, observables which measure, e.g., how close a 
local particle configuration is to rearranging into a different local 
structure could be expected to display violations of quasi-equilibrium 
behavior. If the density of such configurations decreases with
increasing $\tw$ then the bound of~\cite{CugDeaKur97} is essentially void
as discussed below Eq.~(\ref{equ:V}). It would be interesting to construct 
such observables explicitly -- thresholding of an appropriately defined 
free volume would seem an obvious candidate -- and to test our hypothesis 
in simulations. 

Finally, the requirement that the short-time relaxation should display 
quasi-equilibrium behavior could be used to narrow down the class of 
``neutral observables'' which are suitable for measuring a
well-defined effective temperature in the limit of {\em large} time
differences. We note in this context that the condition
$\lim_{\tw\to\infty} (\partial/\partial\tw)C(t,\tw) \neq 0$ at fixed 
$\dt$ is {\em not} necessary to obtain quasi-equilibrium FDT. For the 
single-defect observables considered in~\cite{MayBerGarSol03,BuhGar02b,Buhot03}, 
for instance, this limit vanishes yet quasi-equilibrium behavior is
observed nevertheless.

\begin{acknowledgments} 
We acknowledge financial support from \"{O}ster\-reich\-ische Akademie der 
Wissenschaften and EPSRC Grant No.\ 00800822. 
\end{acknowledgments}

\appendix*
\section{}

We summarize below the ingredients that are needed to obtain our 
expressions for $C \equiv C_1$ and $R \equiv R_1$ based on the 
general results derived in~\cite{MaySol04}. The correlation 
$C(t,\tw)$ is first reduced to multispin correlations by 
substituting $n_i = \frac{1}{2} \left( 1 - \sigma_i \sigma_{i+1} \right)$ 
in (\ref{equ:cor2tdef}). This gives 
\begin{eqnarray*}
  16 \, C(t,\tw) & = & 
      + C_{(0,1),(0,1)} + C_{(0,1),(1,2)} - C_{(0,1),(0,2)} \\
  & & + C_{(1,2),(0,1)} + C_{(1,2),(1,2)} - C_{(1,2),(0,2)} \\
  & & - C_{(0,2),(0,1)} - C_{(0,2),(1,2)} + C_{(0,2),(0,2)}, 
\end{eqnarray*}
where 
\begin{eqnarray}
  C_{\ivec,\jvec} &=& 
    \langle \sigma_{i_1}(t) \sigma_{i_2}(t) \sigma_{j_1}(\tw) 
    \sigma_{j_2}(\tw) \rangle \nonumber \\
  & & - \langle \sigma_{i_1}(t) \sigma_{i_2}(t) \rangle 
    \langle \sigma_{j_1}(\tw) \sigma_{j_2}(\tw) \rangle. 
    \label{equ:corij}
\end{eqnarray}
We have omitted the time-arguments in $C_{\ivec,\jvec}$ in order to 
save space. When using the 
symmetries of $C_{\ivec,\jvec}$ under translations, reflections and 
permutations [among the components of $\ivec$ and $\jvec$ but not
between $\ivec$ and $\jvec$] the  
above equation for $C(t,\tw)$ assumes the simpler form 
\begin{eqnarray}
  C(t,\tw) & = & \frac{1}{8} \left[ C_{(0,1),(0,1)} + C_{(1,2),(0,1)} 
    - C_{(0,1),(0,2)} \right. \nonumber \\ 
  & & \left. -C_{(0,2),(0,1)} \right] + \frac{1}{16} C_{(0,2),(0,2)}. 
  \label{equ:cor2taux2}
\end{eqnarray}
Note that while $C(t,\tw)$ can be expressed in terms of 4-spin 
correlations (\ref{equ:corij}) only, this is not possible for $d \geq 2$; 
in the latter case the expressions for $C_d(t,\tw)$ contain 8-spin two-time 
correlations and exact calculations become exceedingly cumbersome.
In full analogy to $C(t,\tw)$ the response $R$ may be decomposed into 
\begin{eqnarray}
  R(t,\tw) & = & \frac{1}{8} \left[ R_{(0,1),(0,1)} + R_{(1,2),(0,1)} 
    - R_{(0,1),(0,2)} \right. \nonumber \\
  & & \left. -R_{(0,2),(0,1)} \right] + \frac{1}{16} R_{(0,2),(0,2)}, 
  \label{equ:res2taux1}
\end{eqnarray}
with 
\begin{equation}
  R_{\ivec,\jvec} = T \left. \frac{\delta \langle \sigma_{i_1}(t) 
  \sigma_{i_2}(t) \rangle}{\delta h_{\jvec}(\tw)} \right|_{h_{\jvec}=0}. 
  \label{equ:resij}
\end{equation}
For the multispin response function (\ref{equ:resij}) the field
$h_{\jvec}$ is thermodynamically conjugate to $\sigma_{j_1}
\sigma_{j_2}$. Let us remark that while the pair $C$, $R$ violates 
quasi-equilibrium this is not the case~\footnote{In the short-time regime 
the leading contributions of the spin-functions, which satisfy 
quasi-equilibrium~\cite{MayBerGarSol03}, cancel in (\ref{equ:cor2taux2}), 
(\ref{equ:res2taux1}). At fixed $\dt \geq 0$ we have, e.g., $C_{\ivec,\jvec} = 
O\big({\tw}^{-1/2}\big)$ but $C = O\big( {\tw}^{-3/2} \big)$.} 
for the constituting pairs $C_{\ivec,\jvec}$, $R_{\ivec,\jvec}$.
Next, the 4-spin correlations in (\ref{equ:cor2taux2}) are expressed 
in terms of the result given in~\cite{MaySol04}, viz., 
\begin{eqnarray}
  C_{\ivec,\jvec} & = & 
  + \big[ \mathcal{F}_{i_1,i_2}^{\,\jvec} - \Hn_{i_2-i_1}(2\dt,2\tw) \big] \, 
  \Hn_{j_2-j_1}(2\tw) 
  \nonumber \\
  & & 
  - \big[+\me^{-\dt} \, \In_{i_1-j_1}(\gamma \dt) + 
    \mathcal{E}_{i_1,j_1}^{\,\jvec} \big] 
    \nonumber \\
  & & 
  \times \big[-\me^{-\dt} \, \In_{i_2-j_2}(\gamma \dt) + 
    \mathcal{E}_{i_2,j_2}^{\,\jvec} \big] 
  \nonumber \\
  & & 
  + \big[-\me^{-\dt} \, \In_{i_1-j_2}(\gamma \dt) + 
    \mathcal{E}_{i_1,j_2}^{\,\jvec} \big] 
  \nonumber \\
  & & 
  \times \big[+\me^{-\dt} \, \In_{i_2-j_1}(\gamma \dt) + 
    \mathcal{E}_{i_2,j_1}^{\,\jvec} \big]. 
  \label{equ:C22}
\end{eqnarray}
Here and below the indices $\ivec=(i_1,i_2)$ and $\jvec=(j_1,j_2)$ must 
satisfy $i_1<i_2$ and $j_1<j_2$. The multispin response function 
(\ref{equ:resij}) for the case $j_2 = j_1+1$ is also stated explicitly 
in~\cite{MaySol04} and reads [$\jvec^{1,\mathrm{s}} =(j_1-1,j_1)$ and 
$\jvec^{2,\mathrm{s}}=(j_2,j_2+1)$]
\begin{widetext}
\begin{eqnarray}
  R_{\ivec,\jvec} & = & +\me^{-\dt} \, 
  \In_{i_1-j_1} \left[ - \left(1-\frac{\gamma^2}{2} \right)
  \left( - \me^{-\dt} \, \In_{i_2 -j_2} + 
  \mathcal{E}_{i_2,j_2}^{\, \jvec} \right) 
  + 
  \frac{\gamma^2}{2}
  \left( + \me^{-\dt} \, \In_{i_2 -j_1+1} + 
  \mathcal{E}_{i_2,j_1-1}^{\, \jvec^{1,\mathrm{s}}} \right)
  \right] \nonumber \\ 
  & & +\me^{-\dt} \, 
  \In_{i_2-j_1} \left[ + \left(1-\frac{\gamma^2}{2} \right)
  \left( - \me^{-\dt} \, \In_{i_1 -j_2} + 
  \mathcal{E}_{i_1,j_2}^{\, \jvec} \right) 
  - 
  \frac{\gamma^2}{2}
  \left( + \me^{-\dt} \, \In_{i_1 -j_1+1} + 
  \mathcal{E}_{i_1,j_1-1}^{\, \jvec^{1,\mathrm{s}}} \right)
  \right] \nonumber \\ 
  & & -\me^{-\dt} \, 
  \In_{i_1-j_2} \left[ + \left(1-\frac{\gamma^2}{2} \right)
  \left( + \me^{-\dt} \, \In_{i_2 -j_1} + 
  \mathcal{E}_{i_2,j_1}^{\, \jvec} \right) 
  - 
  \frac{\gamma^2}{2}
  \left( - \me^{-\dt} \, \In_{i_2 -j_2-1} + 
  \mathcal{E}_{i_2,j_2+1}^{\, \jvec^{2,\mathrm{s}}} \right)
  \right] \nonumber \\ 
  & & -\me^{-\dt} \, 
  \In_{i_2-j_2} \left[ - \left(1-\frac{\gamma^2}{2} \right)
  \left( + \me^{-\dt} \, \In_{i_1 -j_1} + 
  \mathcal{E}_{i_1,j_1}^{\, \jvec} \right) 
  + 
  \frac{\gamma^2}{2} 
  \left( - \me^{-\dt} \, \In_{i_1 -j_2-1} + 
  \mathcal{E}_{i_1,j_2+1}^{\, \jvec^{2,\mathrm{s}}} \right)
  \right]. 
  \label{equ:R22}
\end{eqnarray}
We have omitted the arguments $(\gamma \dt)$ of the functions $\In$ 
in order to save space. The multispin response functions for the case 
$j_2 > j_1+1$ are not given explicitly in~\cite{MaySol04}. However, by 
following the general procedure developed there one verifies the 
result [$\jvec^{1,\mathrm{s}} =(j_1-1,j_1,j_1+1,j_2)$ and 
$\jvec^{2,\mathrm{s}}=(j_1,j_2-1,j_2,j_2+1)$]
\begin{eqnarray}
  R_{\ivec,\jvec} & = & 
  +\me^{-\dt} \, \In_{i_1-j_1} \bigg\{ 
  \!- \left(1-\frac{\gamma^2}{2} \right)
  \left( - \me^{-\dt} \, \In_{i_2 -j_2} + 
  \mathcal{E}_{i_2,j_2}^{\,\jvec} \right) + \frac{\gamma^2}{2} 
  \Big[
  \left( + \me^{-\dt} \, \In_{i_2 -j_1+1} + 
  \mathcal{E}_{i_2,j_1-1}^{\, \jvec^{1,\mathrm{s}}} \right) \Hn_{j_2-j_1-1} 
  \nonumber \\ 
  & & \qquad 
  - \left( + \me^{-\dt} \, \In_{i_2 -j_1-1} + 
  \mathcal{E}_{i_2,j_1+1}^{\, \jvec^{1,\mathrm{s}}} \right) 
  \Hn_{j_2-j_1+1}
  + \left( - \me^{-\dt} \, \In_{i_2 -j_2} + 
  \mathcal{E}_{i_2,j_2}^{\, \jvec^{1,\mathrm{s}}} \right) 
  \Hn_2
  \Big] \bigg\}
  \nonumber \\ 
  & & +\me^{-\dt} \, 
  \In_{i_2-j_1} \bigg\{ \!+ \left(1-\frac{\gamma^2}{2} \right)
  \left( - \me^{-\dt} \, \In_{i_1 -j_2} + 
  \mathcal{E}_{i_1,j_2}^{\,\jvec} \right) -\frac{\gamma^2}{2}
  \Big[
  \left( + \me^{-\dt} \, \In_{i_1 -j_1+1} + 
  \mathcal{E}_{i_1,j_1-1}^{\, \jvec^{1,\mathrm{s}}} \right) \Hn_{j_2-j_1-1}
  \nonumber \\ 
  & & \qquad 
  - \left( + \me^{-\dt} \, \In_{i_1 -j_1-1} + 
  \mathcal{E}_{i_1,j_1+1}^{\, \jvec^{1,\mathrm{s}}} \right) \Hn_{j_2-j_1+1}
  + \left( - \me^{-\dt} \, \In_{i_1 -j_2} + 
  \mathcal{E}_{i_1,j_2}^{\, \jvec^{1,\mathrm{s}}} \right) \Hn_2
  \Big] \bigg\}
  \nonumber \\ 
  & & -\me^{-\dt} \, 
  \In_{i_1-j_2} \bigg\{ \!+ \left(1-\frac{\gamma^2}{2} \right)
  \left( + \me^{-\dt} \, \In_{i_2 -j_1} + 
  \mathcal{E}_{i_2,j_1}^{\,\jvec} \right) - \frac{\gamma^2}{2}
  \Big[
  \left( + \me^{-\dt} \, \In_{i_2 -j_1} + 
  \mathcal{E}_{i_2,j_1}^{\, \jvec^{2,\mathrm{s}}} \right) \Hn_2
  \nonumber \\ 
  & & \qquad 
  - \left( - \me^{-\dt} \, \In_{i_2 -j_2+1} + 
  \mathcal{E}_{i_2,j_2-1}^{\, \jvec^{2,\mathrm{s}}} \right) \Hn_{j_2-j_1+1}
  + \left( - \me^{-\dt} \, \In_{i_2 -j_1-1} + 
  \mathcal{E}_{i_2,j_2+1}^{\, \jvec^{2,\mathrm{s}}} \right) \Hn_{j_2-j_1-1}
  \Big] \bigg\}
  \nonumber \\ 
  & & -\me^{-\dt} \, 
  \In_{i_2-j_2} \bigg\{ \!- \left(1-\frac{\gamma^2}{2} \right)
  \left( + \me^{-\dt} \, \In_{i_1 -j_1} + 
  \mathcal{E}_{i_1,j_1}^{\,\jvec} \right) + \frac{\gamma^2}{2} 
  \Big[
  \left( + \me^{-\dt} \, \In_{i_1 -j_1} + 
  \mathcal{E}_{i_1,j_1}^{\, \jvec^{2,\mathrm{s}}} \right) \Hn_2
  \nonumber \\
  & & \qquad 
  - \left( - \me^{-\dt} \, \In_{i_1 -j_2+1} + 
  \mathcal{E}_{i_1,j_2-1}^{\, \jvec^{2,\mathrm{s}}} \right) \Hn_{j_2-j_1+1}
  + \left( - \me^{-\dt} \, \In_{i_1 -j_2-1} + 
  \mathcal{E}_{i_1,j_2+1}^{\, \jvec^{2,\mathrm{s}}} \right) \Hn_{j_2-j_1-1}
  \Big] \bigg\}.
  \label{equ:R22big}
\end{eqnarray}
\end{widetext}
Again, all functions $\In$ have argument $(\gamma \dt)$ and additionally 
all functions $\Hn$ appearing in (\ref{equ:R22big}) have argument $(2\tw)$. 
After substituting (\ref{equ:C22}), (\ref{equ:R22}),
(\ref{equ:R22big}) for the multispin correlation and response
functions in (\ref{equ:cor2taux2}) and (\ref{equ:res2taux1}), 
$C(t,\tw)$ and $R(t,\tw)$ are expressed in
terms of $\In, \Hn, \mathcal{E}$ and $\mathcal{F}$. Next we also represent 
$\mathcal{E}$ and $\mathcal{F}$ in terms of $\In$ and $\Hn$ by applying 
the corresponding formulas derived in~\cite{MaySol04}, viz.\ 
\begin{eqnarray*}
  \mathcal{E}_{i_\eps,j_\nu}^{\,\jvec} & = & 
  \Hn_{j_\nu-i_\eps}(\dt,2\tw) \\
  & & - 
  \sum_m \boldsymbol{\delta}_{\jvec,m} \, 
  \me^{-\dt} \, \In_{i_\eps-m}(\gamma \dt) \,
  \Hn_{j_\nu-m}(2\tw), 
\end{eqnarray*}
and 
\begin{eqnarray*}
  \mathcal{F}_{i_\eps,i_\delta}^{\,\jvec} & = & 
  \Hn_{i_\delta-i_\eps}(2\dt,2\tw) \\ 
  & & - \sum_{m} \boldsymbol{\delta}_{\jvec,m} \, 
    \me^{-\dt} \In_{i_\eps-m}(\gamma \dt) \, \Hn_{i_\delta-m}(\dt,2\tw) \\
  & & + \sum_{m} \boldsymbol{\delta}_{\jvec,m} \, 
    \me^{-\dt} \In_{i_\delta-m}(\gamma \dt) \, \Hn_{i_\eps-m}(\dt,2\tw) \\
  & & + \sum_{m,n} \boldsymbol{\delta}_{\jvec,m} \, 
  \boldsymbol{\delta}_{\jvec,n} \, 
  \me^{-2 \dt} \, \In_{i_\eps-m}(\gamma \dt) \, \In_{i_\delta-n}(\gamma \dt) \\
  & & \times \Hn_{n-m}(2\tw),
\end{eqnarray*}
where $\boldsymbol{\delta}_{\jvec,m} = 1- \sgn(j_1-m) \cdots \sgn(j_p-m)$ 
with $p=\mathrm{dim}(\jvec)$, i.e.\ $p=2$ or 4 in our case. The sums in 
these equations are finite since $\boldsymbol{\delta}_{\jvec,m}$ 
is nonzero only within the index-range covered by the components of 
$\jvec$. Upon substitution of the latter equations for all functions 
$\mathcal{E}$, $\mathcal{F}$ the defect-pair correlation and response 
functions are expressed purely in terms of $\In, \Hn$. The functions 
$H$, in turn, are expressed in terms of modified Bessel functions via
\begin{equation}
  \Hn_n(t_1,t_2)=\frac{\gamma}{2} \int_{t_1}^{t_1+t_2} 
  \dd\tau \, \me^{-\tau} \left[ \In_{n-1}-\In_{n+1}\right](\gamma\tau), 
  \label{equ:Hn}
\end{equation}
where we use the notation $\Hn_n(\tau)=\Hn_n(0,\tau)$. In equilibrium the 
quantity $\Hn_{n,\eq}(\tau) = \lim_{t\to\infty} \Hn_n(\tau,t)$ is relevant,
cf.\ Eq.~\ref{equ:coreq}. Simplification of the expressions for $C$ and $R$ 
are possible when using the recursion formula
\begin{eqnarray}
  \Hn_{n+1}(t_1,t_2) & = & -\Hn_{n-1}(t_1,t_2) + \frac{2}{\gamma} \, 
  \Hn_n(t_1,t_2) \nonumber \\
  & & + \me^{-t_1-t_2} \left[ \In_{n-1}-\In_{n+1} \right]\left( \gamma (t_1+t_2) 
  \right) \nonumber \\[1ex]
  & & - \me^{-t_1} \left[ \In_{n-1}-\In_{n+1} \right]\left( \gamma t_1 \right). 
  \label{equ:Hrec}
\end{eqnarray}
One easily proves (\ref{equ:Hrec}) when substituting (\ref{equ:Hn})
and integrating by parts. With $\Hn_0(t_1,t_2)=0$, which follows trivially 
from (\ref{equ:Hn}), any function $\Hn_n(t_1,t_2)$ may thus be decomposed into
modified Bessel functions and $\Hn_1(t_1,t_2)$. 
Also, the recursion $[\In_{n-1}-\In_{n+1}](x) = 
\frac{2n}{x} \In_n(x)$ is useful for rearranging the results.
We use {\tt Mathematica 5.0} to carry out the algebraic manipulations
described above. The procedure yields significant cancellations
in the expressions for $C$ and $R$. For a quench to $T > 0$ we 
obtain (\ref{equ:cor2t}), see below, and a similar expression for $R$; taking 
$T \to 0$ where the integral $H_1$ is soluble~\cite{MaySol04} then 
produces the results (\ref{equ:Csa})-(\ref{equ:Rag}). 
\begin{widetext}
\begin{eqnarray}
  C(t,\tw) & \!\! = \!\! & -\frac{1}{4\gamma^2} \bigg\{ 
  \me^{-(t+\tw)} \frac{\In_1(\gamma(t+\tw))}{t+\tw} \bigg[ 
  \me^{-(t+\tw)} \frac{\In_1(\gamma(t+\tw))}{t+\tw} 
  -2\gamma \me^{-(t-\tw)} [\In_0-\In_1](\gamma(t-\tw)) [1-\Hn_1(2\tw)] \bigg] 
  \nonumber \\
  & & + \Hn_1(t-\tw,2\tw) \bigg[ \frac{2}{\taueq} \me^{-(t+\tw)} 
  \frac{\In_1(\gamma(t+\tw))}{t+\tw} 
  -2\gamma \me^{-(t-\tw)} [\In_0-\In_1](\gamma(t-\tw)) \bigg( 
  \frac{1}{\taueq} + \me^{-2\tw} \frac{\In_1(2\gamma\tw)}{2\tw} \bigg) \bigg] 
  \nonumber \\
  & & + \frac{1}{{\taueq}^2} \Hn_1^2(t-\tw,2\tw) \bigg\}. 
  \label{equ:cor2t}
\end{eqnarray}
\end{widetext}

\bibliography{QuasiEquilibrium}

\end{document}